\documentstyle[12pt]{article}
\def\l{\lambda}
\def\m{\mu}
\def\n{\nu}
\def\r{\rho}
\def\o{\omega}
\def\d{\delta}
\def\e{\epsilon}

\def\be{\begin{equation}}
\def\ee{\end{equation}}
\def\p{\partial}
\def\ber{\begin{eqnarray}}
\def\eer{\end{eqnarray}}
\begin{document}

\begin{center}
{\Large\bf  Gauge Theories on A(dS) Space and Killing Vectors }
\vskip 1 true cm
{\bf Rabin 
Banerjee}\footnote{rabin@bose.res.in}
   and   
{\bf  Bibhas Ranjan Majhi}\footnote{bibhas@bose.res.in}
\vskip .8 true cm
S.N. Bose National Centre for Basic Sciences,\\
 Sector-3, Block-JD, Salt Lake City,  Kolkata 700098, India.\end{center}
\bigskip

\centerline{\large\bf Abstract}
\medskip

We provide a general technique for collectively analysing a manifestly covariant formulation of non-abelian  gauge
theories on both  anti de Sitter  as well as de Sitter spaces. This is done by stereographically projecting the
corresponding theories, 
defined on a flat Minkowski space, onto the
surface of the A(dS)  hyperboloid. The gauge and matter fields in the two descriptions are mapped by conformal Killing vectors and conformal
Killing spinors, respectively. A bilinear map connecting the spinors with the vector is established. Different forms of gauge fixing conditions and their equivalence are discussed. The U(1) axial anomaly as well as the non-abelian covariant and consistent chiral anomalies on A(dS) space are obtained. Electric-magnetic duality is demonstrated. The zero curvature limit is shown to yield consistent findings.

\newpage

\section{Introduction}
 
\bigskip

Quantum field theories on anti de Sitter and  de Sitter, collectively  denoted as A(dS), space-times have a long history originating from the pioneering paper by Dirac \cite{dirac}. These space-times are crucial in cosmological studies since they are the only maximally symmetric examples of a curved space-time manifold. Incidentally, the A(dS) space-time is a solution of the negative (positive) cosmological Einstein's equations having the same degree of symmetry as the flat Minkowski space-time solution. Moreover, recently a non-zero cosmological constant has been proposed to explain the luminosity observations of the farthest supernovae \cite{super}. The A(dS) metric is therefore expected to play an important role if this proposal is validated. These developments show that the study of field theories on A(dS) space is highly desirable, if not essential.

        Most of the available treatments of quantum field theory in curved space-time employ high powered mathematical tools \cite{wald}. While exploiting such techniques for the A(dS) case are feasible, it is not particularly practical since it misses the special symmetry properties of this space-time. Examples of such approaches are the dimensional reduction scheme using vierbein language \cite{siegel,wally} or those based on group theoretical notions \cite{metsaev,FPT}. Yet another method is to use the coordinate independent approach, also called the ambient formalism. For scalar fields this was done in \cite{bros,bros1} which was later extended to include gauge theories \cite{takook,takook1}. While an advantage of this approach is its link (although not a complete one to one mapping) with the corresponding analysis on a flat Minkowski space-time, there is an unpleasant feature which  also exists in other approaches \cite{dirac,metsaev,FPT,bros,bros1,takook,takook1}. The point is that whereas the electron wave equation involves the angular momentum operator, the gauge field equation involves both this operator as well as the ordinary momentum operator. Since the A(dS) space is a hyperboloid (pseudosphere) the natural operator entering into the equation of motion should be the relevant angular momentum operator, since translations on the A(dS) space correspond to rotations on the pseudosphere. This is usually corrected by imposing subsidiary conditions to avoid going off the pseudosphere of constant length.

    In this paper we develop a manifestly covariant method of formulating interacting gauge theories on the A(dS) space-time. Some basic features of this method were already discussed by one of us \cite{banerjee2} in the context of de-Sitter space and its Wick rotated version, the hypersphere which is the n-dimensional sphere immersed in (n+1)-dimensional flat space \cite{banerjee,banerjee1}. The relevant wave operators always incur the angular momentum operators so that subsidiary conditions necessary in other approaches are avoided. The method is general enough to collectively dicuss both the de-Sitter as well as the anti de-Sitter examples. Extention to arbitrary dimensions is straightforward. Our method is applicable for higher rank tensor fields. An exact one to one correspondence with the theories on the flat space has been established. Effectively, the theories on the flat space are projected on to the A(dS) space by a stereographic transformation which is basically a conformal transformation. We show that variables in the gauge sector (like potentials, field strengths etc.) in the two descriptions are related by rules similar to usual tensor analysis, with the conformal Killing vectors playing the role of the metric. Likewise, quantities in the matter (fermionic) sector are related by the conformal Killing spinors. A bilinear map connects these spinors with the conformal Killing vector. Apart from formulating gauge theories we have also computed the chiral anomalies and exhibited electric-magnetic duality rotations in A(dS) space.

  The analysis presented in this paper is basically classical and the extension to quantum field theory will be quite nontrivial. While certain related points are studied in section-5, fully addressing this issue is beyond the scope of the present paper.

       In section-2 the connection between stereographic projection and conformal Killing vectors is shown including an explicit derivation of the latter using the Cartan-Killing equation. The use of these Killing vectors is also elaborated. The pure Yang-Mills theory on A(dS) space is formulated in details in section-3. The action is derived. Its equivalence with the standard action defined on an arbitrary curved space is shown by using the explicit form of the induced metric. The gauge symmetry is discussed and its connection with the gauge identity is shown. Different forms of the Lorentz gauge condition and their equivalence are analysed. Finally, implications of subsidiary conditions used in the literature \cite{dirac,metsaev,FPT,bros,bros1,takook,takook1} are mentioned. Section-4 discusses the stereographic projection of the Dirac lagrangian by means of conformal Killing spinors. The bilinear map connecting these spinors with the conformal Killing vector is given. Section-5 provides a detailed calculation of both the U(1) (axial) anomaly as well as the non-abelian (covariant and consistent) chiral anomalies. The counterterm connecting the covariant and consistent anomaly is also computed. Electric-magnetic duality rotations in an abelian theory are discussed in section-6. A second rank antisymmetric tensor gauge theory is formulated in section-7. The zero curvature limit, analysed in section-8, yields consistent results. The equations of motion on A(dS) space smoothly pass to corresponding equations on the flat Minkowski space. Finally, our conclusions are given in section-9. An appendix discussing the role of boundary conditions has also been included.

\section{ Stereographic Projection and Killing Vectors On A(dS) Space}

Amongst curved spacetimes, the de Sitter and anti-de Sitter spaces are the only possibilities that have
maximal symmetry admitting the highest possible number of Killing vectors. The role of these vectors
in suitably defining gauge theories on such spaces is crucial to this analysis. We shall do our 
discussions for de Sitter and Anti-de Sitter spaces collectively. 

   The A(dS) universe is a pseudosphere in a five dimensional flat space with Cartesian coordinates
$r^a = (r^0, r^1, r^2, r^3, r^4)$ satisfying,
\ber
r^2 &=& r_a r^a 
\nonumber
           \\&=& \eta_{\mu\nu} r^\mu r^\nu+s(r^4)^2 = sl^2  
\label{ds1}
\eer
where $s=-1$ for de Sitter space , $s=+1$ for anti-de Sitter space 
and $l$ is the A(dS) length parameter.
The metric of the de Sitter space $dS(4, 1)$ is induced from the pseudo Euclidean metric
$\eta = diag(+1, -1, -1, -1, -1)$. It has the pseudo orthogonal group $SO(4, 1)$ as the 
group of motions. Anti-de Sitter comes from 
$\eta = diag(+1, -1, -1, -1, +1)$. It has a pseudo orthogonal group $SO(3, 2)$as the group motions.The mostly negative flat Minkowski metric is                        $\eta_{\mu\nu}=diag(+1, -1, -1, -1)$with $\mu, \nu = 0, 1, 2, 3$.

    It should be mentioned that A(dS) is only locally Weyl (conformally) equivalent to the flat space. Although this is enough for our purposes, we remark that the correspondence is globally more intricate and not discussed here. For instance, equation (\ref{ds1}) defines a geometry that can be considered to describe a patch covering only a finite time interval in A(dS); it cannot be used to describe field configurations in the full A(dS) geometry.

 A useful parametrisation of these spaces is done by exploiting the strereographic 
projection. The four dimensional stereographic coordinates $(x^\mu)$ are obtained
by projecting the A(dS) surface into a target Minkowski space. The relevant
equations are \cite{gursey},
\be
r^\mu = \Omega(x) x^\mu \,\,\,;\,\,\, \Omega(x)=\Big(1+s{x^2\over 4l^2}\Big)^{-1}
\label{ds3}
\ee
and,
\be
r^{'4} = -\Omega(x) (1-s{x^2\over 4l^2})
\ee
\label{ds4}
where $x^2 = \eta_{\mu\nu} x^\mu x^\nu$ and $r^{'4} = \frac{r^4}{l} = s\frac{r_4}{l}$.

  The inverse transformation is given by,
\be
x^\mu = \frac{2}{1-r^{'4}} r^\mu
\label{ds5}
\ee

In order to define a gauge theory on the A(dS) space analogous stereographic
projections for gauge fields have to be obtained. This is done following the method
developed by us \cite{banerjee, banerjee1} in the example of the hypersphere which was later extended to the de-Sitter hyperboloid \cite{banerjee2}. The point is that there 
is a mapping of symmetries on the flat space and the pseudosphere (e.g.translations on 
the former are rotations on the latter) that is captured by the relevant Killing vectors.
Furthermore since stereographic projection is known to be a conformal transformation, one
expects that the cherished map among gauge fields would be provided by the conformal Killing
vectors.  We may write this relation as,
 \be
 \hat A_a = K_a^\m A_\m +r_a\phi
 \label{s7}
 \ee
 where the conformal Killing vectors $K_a^\mu$ satisfy the transversality condition,
 \be
 r^a  K_a^\m = 0
 \label{s8}
 \ee
  and an additional scalar field $\phi$, which is just the normal component of $\hat A_a$, {\footnote{
Hat variables are defined on the A(dS) universe while the normal ones
are on the flat space}} is introduced,
 \be
 \phi= s\frac{1}{l^2}r^a\hat A_a
 \label{s9}
 \ee 

The five components of $\hat A$ are expressed in terms of the four components of $A$ plus a scalar degree of freedom. To simplify the analysis the scalar field is put to zero. It is straightforward to resurrect it by using the above equations. With the scalar field gone, $\hat A$ is now given by,
\be
\hat A_a = K_a^\m A_\m
\label{k}
\ee
and satisfies the transversality condition originally used by Dirac \cite{dirac}
 \be
r^a \hat A_a = 0
\label{s6}
\ee

The conformal Killing vectors $K_a^\m$ are now determined. These should satisfy the Cartan-Killing equation which, specialised to a flat four-dimensional manifold, is given by,
\be
\p^\n K_a^\m +\p^\m K_a^\n = \frac{2}{4} \p_\l K_a^\l  \eta^{\m\n}
\label{s10}
\ee

The most general solution for this equation  is given by \cite{FMS},
\be
K_a^\m= t_a^\m +\e_a x^\m + \omega_{a}^{\m\n} x_\n +\l_a^\m x^2 -2\lambda_a^ \sigma x_\sigma x^\m
\label{s11}
\ee
where $\o^{\m\n}= - \o^{\n\m}$.
The various transformations of the conformal group are characterised by the parameters appearing in the above equation; translations by $t$, dilitations by $\e$, rotations by $\o$ and inversions (or the special conformal transformations) by $\l$. Imposing the condition (\ref{s8}) and equating coefficients of terms with distinct powers of $x$, we find the basic structures of the Killing vectors:
\be
K_\n^\m = \Big(1+s\frac{x^2}{4l^2}\Big) \eta_\n^\m - s\frac{x_\n x^\m}{2l^2} 
\label{s18a}
\ee
\be
K_{4}^\m = sK^{4\m} = \frac{x^\m}{l}
\label{s18b}
\ee
These Killing vectors establish the link between the A(dS) coordinates and the flat ones by the relation,
\be
K{_a}{_\m} =  \Big(1+s\frac{x^2}{4l^2}\Big)^2 \frac {\partial{r_a}} {\partial{x^\m}}
\label{deser}
\ee 
With the above solution for the Killing vectors, the stereographic projection for the gauge fields (\ref{k})
is completed leading to, in  component form,
\ber
\hat A_\m &=& \Big(1+s\frac{x^2}{4l^2}\Big) A_\m - s\frac{x^\n x_\m}{2l^2} A_\n\\
\hat A_4 &=& \frac{x_\m}{l} A^\m
\label{newmap}
\eer

 The inverse relation is given by,
\be
\Big(1+s\frac{x^2}{4l^2}\Big)A_\mu = \hat A_\mu + s\frac{x_\mu \hat A_4}{2l}
\label{inverse}
\ee

Before proceeding to discuss gauge theories some properties of these Killing vectors
are summarised. 
There are two useful relations,
\be
K_a^\m K^{a\n} = \Big(1 + s\frac{x^2}{4l^2}\Big)^2 \eta^{\m\n}
\label{k1}
\ee
and,
\ber
K_a^\m K_{b\m} &=&  \Big(1+s\frac{x^2}{4l^2}\Big)^{2}\theta_{ab} 
\nonumber
\\&=& \Big(1+s\frac{x^2}{4l^2}\Big)^{2} \Big(\eta_{ab} - s\frac{r_a r_b}{l^2}\Big)
\label{k2}
\eer
These  relations are also valid for general D-dimensions.
The transverse projector $\theta_{ab}$ satisfies,
\be
r^a \theta_{ab} = r^b \theta_{ab} = 0,\,\,\,\,;\,\,\, \theta_{ab}\theta^{bc}=\theta{_a}{^c}
\ee
and will be subsequently used in the construction of transverse entities like transverse derivatives or angular momentum operators.

              Relation (\ref{k1}) shows that the product of the Killing vectors with repeated `a' indices yields, up to the conformal factor, the induced metric.  The other relation can be interpreted as the transversality condition emanating from (\ref{s8}).
For computing derivatives involving Killing vectors, a particularly useful identity is given by,
\be
K_a^\mu\p_\mu K^{a\nu} = -\Big(1+ s\frac{x^2}{4l^2}\Big) s\frac{x^\nu}{l^2}
\label{k3}
\ee
A simple use of (\ref{k1}) yields the inverse of (\ref{k}) as,
\be
A_\mu = (1+s\frac{x^2}{4l^2})^{-2} K_{a\mu} \hat{A}^a = \frac{\partial{r_a}}{\partial{x^\mu}} \hat{A}^a
\label{bibhas9}
\ee
where the second equality follows from (\ref{deser}). The above relation is an illuminating rephrasing of (\ref{inverse}).

        Note that the conformal (Jacobian) factor that relates the volume element on theA(dS) space 
with that in the four-dimensional flat manifold,
\be
d^4 x = dx_0 dx_1 dx_2 dx_3 = \Big(1+s\frac{x^2}{4l^2}\Big)^4 d\Omega 
\label{conformal}
\ee
naturally emerges in (\ref{k1}) and (\ref{k2}). For D-dimensions this generalises to,
\be
d^D x = dx_0 dx_1 dx_2...... dx_{D-1} = \Big(1+s\frac{x^2}{4l^2}\Big)^D d\Omega 
\ee
 The invariant measure is given by,
\be
d\Omega = \frac{l}{r_4} d^4r =  \Big(\frac{l}{r_4}\Big) dr_0 dr_1 dr_2 dr_3
\label{invmea}
\ee.

      To observe the use of these Killing vectors, 
let us analyse the generators of the infinitesimal
A(dS) transformations. In terms of the host space Cartesian coordinates $r^a$, these
are written as,
\be
L_{ab}=r_a \frac{\p}{\p r^b}- r_b \frac{\p}{\p r^a}
\label{angmom}
\ee
which satisfy the algebra,
\be
[L_{ab}, L_{cd}] = \eta_{bc} L_{ad} + \eta_{ad} L_{bc} - \eta_{bd} L_{ac} - \eta_{ac} L_{bd}
\label{angmomalg}
\ee

In terms of the stereographic coordinates the generator is expressed as,
\be
L_{ab}= (r_a K_b^\mu - r_b K_a^\mu)\p_\mu \,\,\,;\,\,\, \p_\mu = \frac{\p}{\p x^\mu}
\label{v8}
\ee
which can be put in a more illuminating form by contracting with $r^a$,
\be
r^a L_{ab} = sl^2 K_b^\mu\p_\mu
\label{new}
\ee
clearly showing how rotations on the A(dS) space are connected with the translations 
on the flat space by the Killing vectors.

           An object of related interest is the transverse (or tangent) derivative, expressed in terms of the transverse projector,
\be
\nabla^a = \theta^{ab} \frac{\partial}{\partial{r^b}}
\label{bibhas14}
\ee
This derivative satisfies the properties $r_a \nabla^a = 0$ , $\nabla^a r_a = 4$ (or `D' in D-dimensions) and obeys the following commutation relations,
\be
[\nabla^a , \nabla^b] = s\frac{1}{l^2} L^{ab}\,\,\,;\,\,\, [\nabla^a , r^b] = \theta^{ab}
\label{sc6}
\ee
Also, the angular momentum operator is directly written in terms of the transverse derivative as,
\be
L^{ab} = r^a \nabla^b - r^b \nabla^a\,\,\,;\,\,\, \nabla^a = K^{a\mu}\partial_\mu = s\frac{1}{l^2}r_b L^{ba}
\label{bibhas13}
\ee
where we have used (\ref{new}).

\section{Formulation of Yang-Mills theory on A(dS) space}
\bigskip

In this section we discuss the formulation of Yang-Mills theory on the A(dS) space.
The theory is obtained by stereographically projecting the usual theory defined on the flat Minkowski
space. A comparison with other approaches will be done pointing out the advantages of our formalism.

   The pure Yang-Mills theory on the Minkowski space is governed by the standard Lagrangian,
\be
{\cal L}= -{1\over 4} Tr(F_{\mu\nu} F^{\mu\nu})
\label{v1}
\ee
where the field tensor is given by,
\be
F_{\mu\nu}=\p_\mu A_\nu - \p_\nu A_\mu -i[ A_\mu,  A_\nu ]
\label{v2}
\ee
 
 To define the field tensor on the A(dS) space we proceed systematically by looking at
the gauge symmetries.
If the ordinary potential transforms as,
\be
A_\mu' = U^{-1} (A_\mu +i\p_\mu )U
\label{gt1}
\ee
then the projected potential transforms as,
\be
\hat A_a' = K_a^\m A_\mu'=  U^{-1} (\hat A_a +s\frac{i}{l^2} r^b L_{ba} )U
\label{gt2}
\ee
obtained by using (\ref{k}) and (\ref{new}).

The infinitesimal version of these transformations obtained by taking $U=e^{-i\lambda}$ is then
found to be,
\be
\delta A_\mu = D_\mu\lambda = \p_\mu \lambda -i[ A_\mu, \lambda ]
\label{v5}
\ee
 for the flat space while for the A(dS) space it is given by,
\be
\delta \hat A_a = K{_a}{^\mu}\delta A_\mu=s\frac{1}{l^2}r^b L_{ba}\lambda -i[ \hat A_a , \lambda]
\label{v10}
\ee
This is put in a more transparent form by introducing, in analogy with the flat space,
a `covariantised angular derivative' \cite{banerjee1, jackiw} on the A(dS) space,
\be
\hat{\cal L}_{ab} = L_{ab} - i [r_a \hat A_b - r_b \hat A_a,\,\,\, ] = -\hat{\cal L}_{ba} 
\label{new1}
\ee
so that,
\be
\delta \hat A_a = s\frac{1}{l^2}r^b \hat{\cal L}_{ba}\lambda 
\label{new2}
\ee
Note that, this is consistant with $r^a \delta{\hat{A}_a} = 0$ which is a consequence of the transversality condition (\ref{s6}). 

     The covariantised angular derivative satisfies a relation that is the covariantised version
of (\ref{new}),
\be
r^a \hat{\cal L}_{ab} = sl^2 K_b^\mu D_\mu
\label{covnew}
\ee
obtained by using the transversality condition on the gauge fields.

      It is feasible to extend the definition (\ref{bibhas14}) of the transverse derivative to the `covariantised transverse derivative' as,
\be
\tilde{\nabla}^a = \nabla^a -i[\hat{A}^a,]
\label{sc3}
\ee
This is related to the `covariantised angular momentum' (\ref{new1}) by equations similar to (\ref{sc6},\ref{bibhas13}). These are given by their covariantised versions,
\ber
\tilde{\nabla}^a &=& K^{a\mu}D_\mu = K^{a\mu}(\partial_\mu -i[A_\mu,])
\nonumber
\\
\nonumber
\\&=& \frac{s}{l^2}r_b \hat{\cal{L}}^{ba}
\eer
and,
\ber
[\tilde{\nabla}^a,r^b]=\theta^{ab}
\nonumber
\\
\nonumber
\\
\nonumber
[\tilde{\nabla}^a,\tilde{\nabla}^b]=\frac{s}{l^2}\hat{\cal{L}}^{ab}
\eer
Also, $\tilde{\nabla}^a$ satisfies the following properties that are exactly identical to $\nabla^a$;
\ber
r_a\tilde{\nabla}^a = r_a \nabla^a=0
\nonumber
\\
\nonumber
\\\tilde{\nabla}^a r_a =\nabla^a r_a =4
\eer

           The field tensor $\hat F_{abc}$ on the A(dS) space is now defined. It has to be
a fully antisymmetric three index object  that transforms covariantly. The
covariantised angular derivative is the natural choice for constructing it. We define,
\be
\hat F_{abc} = \Big( L_{ab}\hat A_c -ir_a [\hat A_b, \hat A_c]\Big) + c.p.
\label{v12}
\ee
where $c.p.$ stands for the other pair of terms involving cyclic permutations in $a, b, c$.

To see that $\hat{F}_{abc}$ transforms covariantly it is convenient to recast this in a form involving the Killing vectors,
analogous to the relation (\ref{k}). Indeed it is mapped to the field tensor on the flat space by the following
relation,
\be
\hat F_{abc}= \Big(r_a K_b^\mu K_c^\nu +r_b K_c^\mu K_a^\nu + r_c K_a^\mu K_b^\nu\Big)F_{\mu\nu}
\label{v11}
\ee
so that symmetry properties under exchange of the indices is correctly preserved.
To show the equivalence, (\ref{k}) and (\ref{v8}) are used to simplify (\ref{v12}),
yielding,
\be
\hat F_{abc} = \Big(r_a K_b^\mu - r_b K_a^\mu\Big)\p_\mu\Big(K_c^\nu A_\nu\Big)
-i r_a \Big[K_b^\nu A_\nu, K_c^\mu A_\mu \Big] + c.p.
\label{v13}
\ee
The derivatives acting on the Killing vectors sum up to zero on account of the identity,
\be
\Big(r_a K_b^\mu - r_b K_a^\mu\Big)\p_\mu K_c^\nu  +c.p. = 0
\label{v14}
\ee
The derivatives acting on the potentials, together with the other pieces, combine to
reproduce (\ref{v11}), thereby completing the proof of the equivalence.It is now trivial to see, using the above relation (\ref{v11}), that $\hat{F}_{abc}$ transforms
covariantly simply because $F_{\mu\nu}$ does. The exact transformation is given by,
\be
\delta\hat{F}_{abc}=-i[\hat{F}_{abc},\lambda]
\label{tinku1}
\ee
Of course this can also be derived by starting from (\ref{v12}) and using (\ref{v10}).

               The inverse relation following from (\ref{v11}), obtained by contracting with $r^a$ and the Killing vectors, is given by ,
\ber
F_{\mu_\nu}(x) &=& s\frac{1}{l^2}(1+s\frac{x^2}{4l^2})^{-4} K_{b\mu} K_{c\nu} (r_a \hat{F}^{abc}) 
\nonumber
\\
\nonumber
\\&=& s\frac{1}{l^2} \frac{\partial{r}_b}{\partial{x^\mu}} \frac{\partial{r}_c}{\partial{x^\nu}}(r_a \hat{F}^{abc}(r))
\label{bibhas10}
\eer
where use has been made of (\ref{deser}) to get the final result.

         Incidentally the generalisation of (\ref{bibhas9}) and (\ref{bibhas10})to arbitrary rank tensors is easily done, 
\be
A_{\mu_1 \mu_2 ......\mu_n}(x) = \frac{\partial{r}_{a_1}}{\partial{x^{\mu_1}}} \frac{\partial{r}_{a_2}}{\partial{x^{\mu_2}}}.......\frac{\partial{r}_{a_n}}{\partial{x^{\mu_n}}} \hat{A}^{a_1 a_2.....a_n}(r)
\label{bibhas21}
\ee
and
\be
F_{\mu_1 \mu_2 ......\mu_n \mu_{n+1}}(x) = s\frac{1}{l^2} \frac{\partial{r}_{b_1}}{\partial{x^{\mu_1}}} \frac{\partial{r}_{b_2}}{\partial{x^{\mu_2}}}.......\frac{\partial{r}_{b_{n+1}}}{\partial{x^{\mu_{n+1}}}} (r_a \hat{F}^{a b_1 b_2.....b_{n+1}}(r))
\label{bibhas22}
\ee
where $F(\hat{F})$ denotes the field strength corresponding to the potential $A(\hat{A})$.

                    It is possible to extend the above relations to establish a mapping between the covariant derivatives on the A(dS) local coordinates and the transverse derivatives (\ref{bibhas14}) or the angular momentum operator (\ref{bibhas13}). Let us first recall that the  A(dS) space is immersed in a (4+1) dimensional flat space which has the metric $ds^2 = \eta{_a}{_b} dr^a dr^b$ and A(dS) is the subspace $\eta_{ab}r^ar^b = sl^2$ with metric $ds^2 = g{_\mu}{_\nu} dx^\mu dx^\nu$ . So, locally both metrics must agree on A(dS);i.e.                      
\be   
\eta{_a}{_b} dr^a dr^b = g{_\mu}{_\nu} dx^\mu dx^\nu 
\ee
 But,
\ber
 dr^a = \partial{_\mu}r^a dx^\mu &=& \frac{\partial r^a}{\partial x^\mu} dx^\mu 
\nonumber
\\&=& (1+s\frac{x^2}{4l^2})^{-2} K{^a}{_\mu} dx^\mu
\eer   
Hence, 
\be
(1+s\frac{x^2}{4l^2})^{-4} \eta^{ab} K_{a\mu} K_{b\nu} = g{_\mu}{_\nu}
\label{ma1}
\ee 
Using the identity (\ref{k1}), we obtain the induced metric and its inverse,    \be
g{_\mu}{_\nu} = (1+s\frac{x^2}{4l^2})^{-2}\eta{_\mu}{_\nu}\,\,\,;\,\,\, g{^\mu}{^\nu} = (1+s\frac{x^2}{4l^2})^{2}\eta{^\mu}{^\nu}.
\label{new3}
\ee
Incidentally the induced metric $g_{\mu\nu}$ is related to the transverse projector $\theta_{ab}$ (see \ref{k2}) by,
\be
g_{\mu\nu} = \frac{\partial{r^a}}{\partial{x^\mu}}\frac{\partial{r^b}}{\partial{x^\nu}} \theta_{ab}
\ee
which is similar to (\ref{bibhas10}). This follows from (\ref{ma1}), the identification (\ref{deser}) and the transversality condition (\ref{s8}).

The Affine connection for the vector field, 
\be
\Gamma{_\mu}{_\sigma}{^\nu} = \frac{1}{2} g{^\nu}{^\rho}(g{_\rho}{_\mu}{_,}{_\sigma} + g{_\rho}{_\sigma}{_,}{_\mu} - g{_\mu}{_\sigma}{_,}{_\rho})               \ee     
is then found to be,
\be
\Gamma{_\mu}{_\sigma}{^\nu} = s\frac{1}{2l^2}(1+s\frac{x^2}{4l^2})^{-1}(x^\nu \eta_{\mu \sigma} - x_\mu \eta{_\sigma}{^\nu} - x_\sigma \eta{_\mu}{^\nu})
\ee
The covariant derivative acting on a scalar and a covariant vector yields,
\be
\nabla_\mu \phi= \partial_\mu \phi
\ee
and    
\ber
\nabla_\mu A_\nu &=& \partial{_\mu} A_\nu - \Gamma{_\mu}{_\nu}{^\sigma} A_\sigma
\nonumber
\\
\nonumber
\\&=& \partial{_\mu}A_\nu - s\frac{1}{2l^2}(1+s\frac{x^2}{4l^2})^{-1}(x_\sigma A^\sigma \eta{_\mu}{_\nu} - x_\mu A_\nu - x_\nu A_\mu)
\label{field16}
\eer
These results are reproduced by the maps,
\be
\nabla_\mu \phi(x) = \frac{\partial{r_a}}{\partial{x^\mu}} \nabla^a \phi(r)
\label{bibhas11}
\ee
and
\be
\nabla_\mu A_\nu(x)= \frac{\partial{r_a}}{\partial{x^\mu}}\frac{\partial{r_b}}{\partial{x^\nu}}  \nabla^a \hat{A}^b(r)
\label{bibhas12}
\ee
To show this, consider the first of the above relations. Using (\ref{deser}) and the definition (\ref{bibhas13}) of the transverse derivative $\nabla^a$, we obtain,
\be
\frac{\partial{r_a}}{\partial{x^\mu}} \nabla^a \phi = (1+s\frac{x^2}{4l^2})^{-2}K_{a\mu}K^{a\sigma}\partial_\sigma \phi = \partial_\mu \phi
\ee
that follows on using the identity (\ref{k1}). The cherished result $\nabla_\mu \phi = \partial_\mu \phi$ is thereby reproduced. 

    Now using (\ref{deser}), (\ref{k}), (34) and the identity (\ref{k1}), the R.H.S. of (\ref{bibhas12}) becomes,
\be
\frac{\partial{r_a}}{\partial{x^\mu}} \frac{\partial{r_b}}{\partial{x^\nu}} \nabla^a \hat{A}^b(r) = (1+s\frac{x^2}{4l^2})^{-2} K_{b\nu} \partial_\mu (K^{b\sigma} A_\sigma)
\ee
Putting the stuctures of the Killing vectors (\ref{s18a}, \ref{s18b}) above, we get the R.H.S. of  (\ref{field16}), which proves (\ref{bibhas12}).

       Now, it is possible to generalise the relations (\ref{bibhas11},\ref{bibhas12}) to include a chain of covariant derivatives. This leads to the following results,
\be
\nabla_\mu \nabla_\nu \phi(x) = \frac{\partial{r_a}}{\partial{x^\mu}} \frac{\partial{r_b}}{\partial{x^\nu}}\nabla^a \nabla^b \phi(r)
\label{ma5}
\ee
and
\be
\nabla_\mu\nabla_\nu A_\sigma(x)=\frac{\partial{r_a}}{\partial{x^\mu}} \frac{\partial{r_b}}{\partial{x^\nu}}\frac{\partial{r_c}}{\partial{x^\sigma}} \nabla^a\nabla^b \hat{A}^c (r)
\label{ma6}
\ee
The generalisation to higher orders is straight forward. The d'Alembertian on the scalar field is next calculated, using (\ref{ma5}),
\ber
\Box \phi(x) &=& g^{\mu\nu} \nabla_\mu\nabla_\nu \phi(x) 
\nonumber
\\
\nonumber
\\&=&(1+s\frac{x^2}{4l^2})^2\eta^{\mu\nu}\frac{\partial{r_a}}{\partial{x^\mu}}\frac{\partial{r_b}}{\partial{x^\nu}} \nabla^a\nabla^b\phi(r)
\nonumber
\\
\nonumber
\\&=&\nabla^a\nabla_a \phi(r)=\nabla^2\phi(r)
\label{ma3}
\eer
Likewise, the d'Alembertian on the vector potential yields, using (\ref{ma6}),
\be
\Box A_\sigma(x)=g^{\mu\nu}\nabla_\mu\nabla_\nu A_\sigma(x) = \frac{\partial{r_c}}{\partial{x^\sigma}}\nabla^2 \hat{A}^c(r)
\ee

        The action for the Yang-Mills theory on the
A(dS) space is now defined by first considering the repeated product of the field tensors.
Taking (\ref{v11}) and using the
transversality of the Killing vectors, we get,
\be
\hat F_{abc}\hat F^{abc}=  3sl^2\Big(K^{b\mu} K^{c\nu} K_b^\lambda K_c^\rho\Big)F_{\mu\nu} F_{\lambda\rho}
\label{v18}
\ee
Finally, using (\ref{k1}), we obtain,
\be
\hat F_{abc}\hat F^{abc}= 3sl^2 \Big(1+s\frac{x^2}{4l^2}\Big)^4 F_{\mu\nu} F^{\mu\nu}
\label{v19}
\ee
Using this identification as well as (\ref{conformal}),
the actions on the flat space and the A(dS) space are mapped as,
\be
S =  -{1\over 4}\int d^4x Tr.(F_{\mu\nu} F^{\mu\nu}) = -s{1\over {12 l^2}}\int d\Omega Tr.(\hat F_{abc}\hat F^{abc})
\label{v20}
\ee
The lagrangian following from this action is given by,
\be
{\cal L}_\Omega =-s{1\over {12 l^2}}Tr.(\hat F_{abc}\hat F^{abc})
 \label{v21}
\ee
This completes the construction of  the Yang-Mills action which can be taken as the starting point for
 calculations  on the A(dS)
space. This action is manifestly gauge invariant under (\ref{tinku1}). Later on we will discuss an alternative approach to understand this invariance.

      Now using the definition of the induced metric(\ref{new3}), it is possible to show that the above action can also be obtained from the standard action defined on the curved space which is taken as,       
\be
S = - {1\over 4}\int d^4x (\surd{-g}) g{^\mu}{^\rho}g{^\nu}{^\sigma} Tr.(F{^D}{_\mu}{_\nu}F{^D}{_\rho}{_\sigma})
\label{new4}
\ee
where 
\be
F{^D}{_\mu}{_\nu} =  \nabla_\mu A_\nu - \nabla_\nu A_\mu - i[A_\mu , A_\nu]
\ee
and $A_\mu$ has Weyl weight zero for the above action to be conformally invariant.
Hence
\ber
F{^D}{_\mu}{_\nu}F{^D}{_\rho}{_\sigma}
&=& (\partial_\mu A_\nu - \Gamma{_\mu}{_\nu}{^\sigma}A_\sigma - \partial_\nu A_\mu + \Gamma{_\nu}{_\mu}{^\sigma}A_\sigma - i [A_\mu , A_\nu])(\partial_\rho A_\sigma 
\nonumber
\\
\nonumber
\\ &-&\Gamma{_\rho}{_\sigma}{^\lambda}A_\lambda - \partial_\sigma A_\rho + \Gamma{_\sigma}{_\rho}{^\lambda}A_\lambda - i [A_\rho , A_\sigma]) 
\eer
Since $\Gamma{_\mu}{_\nu}{^\sigma}$ is symmetric in two lower indices,
\be
F{^D}{_\mu}{_\nu}F{^D}{_\rho}{_\sigma} = F{_\mu}{_\nu} F{_\rho}{_\sigma}
\ee
where $F{_\mu}{_\nu}$ is defined in (\ref{v2}). Taking the explicit form of the induced metric (\ref{new3}) on the A(dS) space, we obtain,
\ber
\surd(-g) &=& \surd(-det(g{_\mu}{_\nu})) 
\nonumber
\\&=& \surd(-((1+s\frac{x^2}{4l^2})^{-8})det \eta{_\mu}{_\nu})
\nonumber
\\&=& (1+s\frac{x^2}{4l^2})^{-4} 
\label{new7}
\eer
Putting all these values in(\ref{new4}) and using(\ref{conformal}), we get
\be
S = - \frac{1}{4} \int d\Omega (1+s\frac{x^2}{4l^2})^4 (1+s\frac{x^2}{4l^2})^{-4} (1+s\frac{x^2}{4l^2})^4 \eta{^\mu}{^\rho} \eta{^\nu}{^\sigma} Tr.(F{_\mu}{_\nu} F{_\rho}{_\sigma})
\ee
Finally using (\ref{v19}) we obtain 
\be
S  = - s \frac{1}{12l^2}\int d\Omega Tr.(\hat{F}{_a}{_b}{_c} \hat{F}{^a}{^b}{^c})
\label{field9}
\ee
which reproduces (\ref{v20}).

          This section is concluded by providing a brief discussion of the gauge fixing condition. On the hypersphere, Adler \cite{adler} used the following analogue of the Lorentz condition,
\be
L_{ab}\hat{A}^b - \hat{A}_a = 0
\label{laltu31}
\ee
The same condition is also viable on the A(dS) pseudosphere. Since there is a free index in (\ref{laltu31}) its connection with the Lorentz gauge is not particularly transparent. A straightforward algebra, however, yields,
\be
L_{ab}\hat{A}^b - \hat{A}_a = s \frac{1}{l^2}r_a (r_b L^{bc}\hat{A}_c)
\label{ma10}
\ee
so that (\ref{laltu31}) may be equivalently replaced by,
\be
r^aL_{ab}\hat{A}^b = 0
\label{laltu30}
\ee
as the pseudospherical analogue of the Lorentz condition.
In this form there is no free index. Self consistency between (\ref{laltu31}) and (\ref{laltu30}) is established by contracting the former with $r^a$ and using the transversality condition (\ref{s6}).

     In order to prove the identity (\ref{ma10}) we simplify its R.H.S. as follows,
\ber
\frac{s}{l^2} r_a r_b L^{bc}\hat{A}_c&=& r_a K^{c\mu}\partial_\mu \hat{A}_c
\nonumber
\\
\nonumber
\\&=&(r_aK^{c\mu}-r^c K{_a}{^\mu})\partial_\mu \hat{A}_c + r^c K{_a}{^\mu}\partial_\mu\hat{A}_c
\nonumber
\\
\nonumber
\\&=&L{_a}{^c}\hat{A}_c+r^c K{_a}{^\mu}\partial_\mu (K{_c}{^\sigma} A_\sigma)
\eer
where (\ref{v8}) and (\ref{new}) have been used. Now, exploiting the identity,
\be
r^c K{_a}{^\mu}\partial_\mu K{_c}{^\sigma} = -K{_a}{^\sigma}
\label{m12b}
\ee
and the transversality condition (\ref{s6}), we obtain,
\ber
\frac{s}{l^2} r_a r_b L^{bc}\hat{A}_c&=& L{_a}{^c}\hat{A}_c - K{_a}{^\sigma}A_\sigma
\nonumber
\\
\nonumber
\\&=& L{_a}{^c}\hat{A}_c - \hat{A}_a
\eer
thereby proving (\ref{ma10}).

    Actually, it is possible to generalise the relation (\ref{ma10}) to any vector $\hat{V}_a$ which is projected to the flat space by,
\be
\hat{V}_a=(1+s\frac{x^2}{4l^2})^n K{_a}{^\mu}V_\mu
\label{ma11}
\ee
One follows the same steps as before leading to the equivalance,
\be
L_{ab}\hat{V}^b - \hat{V}_a = \frac{s}{l^2}r_a (r_b L^{bc}\hat{V}_c)
\label{ma12}
\ee
This will be used later. Note that, the mapping for the vector potential (\ref{k}) corresponds to putting $n=0$ in (\ref{ma11}).

              It is simple to prove that the operator in (\ref{laltu30}) corresponds to the covariant derivative $\nabla_\mu A^\mu$ in the local coordinates. From (\ref{bibhas12}), we obtain,
\ber
\nabla_\mu A^\mu = g^{\mu\nu}\nabla_\mu A_\nu &=& (1+s\frac{x^2}{4l^2})^2 \eta^{\mu\nu}\frac{\partial{r_a}}{\partial{x^\mu}}\frac{\partial{r_b}}{\partial{x^\nu}} \nabla^a \hat{A}^b 
\nonumber
\\
\nonumber
\\&=& (1+s\frac{x^2}{4l^2})^{-2} K_{a\mu}K{_b}{^\mu} \nabla^a \hat{A}^b 
\eer
Now using the identity (\ref{k2}) and (\ref{bibhas13}) we have,
\be
\nabla_\mu A^\mu = s\frac{1}{l^2} r^b L_{ba} \hat{A}^a
\ee
This further justifies (\ref{laltu30}) as the pseudospherical analogue of the Lorentz gauge.

    The gauge condition in the form (\ref{laltu31}) is particularly useful for simplifying the equation of motion. To see this consider the non-abelian equation of motion obtained from (\ref{field9}), by employing the variational principle {\footnote{See the appendix for details.}},
\be
\hat{\cal{L}}_{ab} \hat{F}^{abc} = 0
\label{field10}
\ee  
where the covariantised angular momentum $\hat{\cal{L}}_{ab}$ is defined in (\ref{new1}). This equation is next written in terms of the potential $\hat{A}_a$. Using the definition of the field tensor (\ref{v12}), the above equation reduces to,
\ber
L_{ab}L^{ab}\hat{A}^c - 2L_{ab}L^{ac}\hat{A}^b-2iL_{ab}\{r^a[\hat{A}^b,\hat{A}^c]\}-iL_{ab}\{r^c[\hat{A}^a,\hat{A}^b]\}
\nonumber
\\
\nonumber
\\-i[r_a\hat{A}_b-r_b\hat{A}_a,\hat{F}^{abc}]=0
\label{field11}
\eer
Now, using (\ref{angmomalg}) we obtain,
\be
[L_{ab},L^{ac}]=-3L{_b}{^c}
\ee
Exploiting this identity in (\ref{field11}) yields,
\ber
L_{ab}L^{ab}\hat{A}^c +6L^{bc}\hat{A}_b-2L^{ac}L_{ab}\hat{A}^b -2iL_{ab}\{r^a[\hat{A}^b,\hat{A}^c]\}-iL_{ab}\{r^c[\hat{A}^a,\hat{A}^b]\}
\nonumber
\\
\nonumber
\\-i[r_a\hat{A}_b-r_b\hat{A}_a,\hat{F}^{abc}]=0
\label{field12}
\eer
In terms of the `covariantised angular momentum' (\ref{new1}) this is written as,
\be
P_{ac}\hat{A}^c = 0
\label{time1}
\ee
where the wave operator $P_{ac}$ is defined by
\be
P_{ac}=\hat{\cal{L}}_{bd}\hat{\cal{L}}^{bd}\eta_{ac}-6\hat{\cal{L}}_{ac}+2\hat{\cal{L}}_{ab}\hat{\cal{L}}{^b}{_c}
\ee
Exploiting the angular momentum algebra (\ref{angmomalg}) it is possible to check the following properties of the wave operator,
\ber
\hat{\cal{L}}^{bc}P_{ca} = P^{bc}\hat{\cal{L}}_{ca} = P{^b}{_a};
\nonumber
\\
\nonumber
\\r^b P_{ba} = P_{ba}r^a = 0
\label{time3}
\eer
It is worthwhile to point out the significance of the identities (\ref{time3}). These are related to the gauge invariance of the action (\ref{field9}) under the transformations (\ref{new2}). The action (\ref{field9}) is manifestly gauge invariant under (\ref{tinku1}). However there is an alternative way of understanding this invariance which illuminates its connection with the gauge identity. As is known, for gauge theories defined on a flat space, gauge invariance of an action is enforced by a gauge identity; the number of gauge parameters being equal to the number of gauge identities. To see this in the present context, we consider the variation of the action (\ref{field9}) under an arbitrary transformation of the potential, $\delta\hat{A}_a$;
\ber
\delta S &=&\int d\Omega Tr.[(\delta\hat{A}_c)(\hat{\cal{L}}_{ab}\hat{F}^{abc})]
\nonumber
\\
\nonumber
\\&=&\int d\Omega Tr.[(\delta\hat{A}_c)(P^{ca}\hat{A}_a)]
\eer
where, for simplicity, we have omitted the prefactor $(\frac{s}{12l^2})$. Invariance of the action under an arbitrary $\delta\hat{A}_a$ yields the equation of motion (\ref{field10}) or (\ref{time1}). If the gauge transformation (\ref{new2}) is now considered, then,
\be
\delta S = \frac{s}{l^2}\int d\Omega Tr.[r^b(\hat{\cal{L}}_{bc}\lambda)(P^{ca}\hat{A}_a)]
\ee
Using an integration by parts we find,
\be
\delta S= -\frac{s}{l^2}\int d\Omega Tr.[\lambda  \hat{\cal{L}}_{bc}\{r^bP^{ca}\hat{A}_a\}]
\ee
Gauge invariance of the action (i.e. $\delta S=0$) requires that the factor multiplying the gauge parameter $\lambda$ should vanish identically, i.e. without recourse to any equations of motion. This indeed happens as a consequence of the properties (\ref{time3}),
\ber
\hat{\cal{L}}_{bc}\{r^bP^{ca}\hat{A}_a\} &=&(\hat{\cal{L}}_{bc}{r^b}) P^{ca}\hat{A}_a+r^b\hat{\cal{L}}_{bc}P^{ca}\hat{A}_a
\nonumber
\\
\nonumber
\\&=&(L_{bc}r^b)P^{ca}\hat{A}_a+r^b P{_b}{^a}\hat{A}_a
\nonumber
\\
\nonumber
\\&=&-(\partial_b r^b)r_c P^{ca}\hat{A}_a=0
\eer
where the definition (\ref{new1}) of the `covariantised angular momentum' $\hat{\cal{L}}_{bc}$ has been explicitly used to simplify the first piece.
The above identity is usually referred as a gauge identity.

       Let us now consider the gauge condition (\ref{laltu31}). It is equivalently expressed as follows,
\ber
\hat{\cal{L}}_{ab} \hat{A}^b - \hat{A}_a &=& L_{ab}\hat{A}^b -i[r_a\hat{A}_b - r_b \hat{A}_a,\hat{A}^b] - \hat{A}^a
\nonumber
\\
\nonumber
\\&=&L_{ab}\hat{A}^b - \hat{A}^a=0
\label{time2}
\eer
where the transversality condition (\ref{s6}) forces the commutator term to vanish. Now using the gauge condition (\ref{time2}) in (\ref{time1}) we obtain,
\be
(\hat{\cal{L}}_{ab}\hat{\cal{L}}^{ab}-4)\hat{A}^c=0
\label{ma4}
\ee

Some comments concerning the equation of motion (\ref{ma4}) are now in order. This equation  dose not involve the parameter `s' and is identical for both AdS as well as dS spaces. Of course this parameter enters when interactions are introduced. For instance, as discussed later, for Yang-Mills field coupled to fermionic matter, the equation of motion (\ref{field10}) is replaced by,
\be
\frac{s}{2l^2}\hat{\cal{L}}_{ab}\hat{F}^{abc}+\hat{j}^c=0
\ee
where $\hat{j}_a$ is the fermionic current. Imposing the gauge condition (\ref{laltu31}) yields the form analogous to (\ref{ma4}),
\be
(\hat{\cal{L}}_{ab}\hat{\cal{L}}^{ab}-4)\hat{A}^c=-2sl^2\hat{j}^c
\label{104.1}
\ee
Finally, note that all derivatives occur only through the angular momentum operator which is the correct derivative operator on the A(dS) pseudosphere.

        The last point mentioned above is usually violated in other formulations \cite{dirac,metsaev,FPT,bros,bros1,takook,takook1} where the wave operator has a rather complicated  structure  so that derivatives involve both $L_{ab}$ as well as $\partial_a$. Hence, to ensure that it dose not go off the pseudosphere, subsidiary conditions are imposed on the field variables. This is now elaborated further.

        In the ambient space formulation, starting from the Casimir operator and using the infinitesimal generators, the Casimir eigenvalue equation and some algebric properties, the field equation for a massless abelian vector field in A(dS) space is found to be \cite{takook} {\footnote{This paper discusses only the de Sitter example while refs. \cite{metsaev,FPT} consider the anti de Sitter case. There is no collective discussion.}},
\be
(sl^2\nabla^2-2)\hat{A}_a +2\frac{s}{l^2}r_ar^bL_{bc}\hat{A}^c-sl^2\nabla_a (\partial^b\hat{A}_b)=0
\label{sc1}
\ee
A similar equation is also given in \cite{metsaev} which exploits the irreducible representations of the kinamatical A(dS) groups or using the triplet formalism \cite{FPT}. The subsidiary (or divergenceless) condition $\partial^b \hat{A}_b =0$ is next imposed to eliminate the last term that explicitly involves the derivative $\partial^b$. For comparing (\ref{sc1}) with our result, it is now necessary to identify $\nabla^2$ with $L_{ab}L^{ab}$. 
 To see this we use the expression for the angular momentum operator (\ref{bibhas13}), to derive,
\be
\frac{s}{2l^2}L_{ab}L^{ab} = \nabla^2 + \frac{s}{l^2}(\nabla^b r^a)(r_a\nabla_b)\ee
where we have used the results, $r_a\nabla^a=0$ and $\nabla^a r_a = 4$. Then,
\ber
\nabla^2 - \frac{s}{2l^2}L_{ab}L^{ab} &=& -\frac{s}{l^2}(\nabla^b r^a)(r_a\nabla_b)
\nonumber
\\
\nonumber
\\&=& (1+s\frac{x^2}{4l^2})^2 r_a (\partial_\mu r^a)\partial^\mu
\nonumber
\\
\nonumber
\\&=&(1+s\frac{x^2}{4l^2})^2 \frac{1}{2}\partial_\mu(r_ar^a)\partial^\mu =0
\label{sc5}
\eer
where the constancy of the A(dS) pseudosphere $r_ar^a=sl^2$ is used to get the vanishing result.
Hence, we obtain the identification,
\be
\nabla^2 = \frac{s}{2l^2}L_{ab}L^{ab}
\label{sc2}
\ee
The Lorentz gauge fixing condition in the form (\ref{laltu30}) is now used to eliminate the second term in (\ref{sc1}). Finally, putting (\ref{sc2}) in (\ref{sc1}) we find,
\be
(L_{bc}L^{bc}-4)\hat{A}_a=0
\ee
which exatly corresponds to the abelian version of (\ref{ma4}). It might be recalled that (\ref{ma4}) was also obtained by using the Lorentz gauge, albeit in the form (\ref{laltu31}). Incidentally it is also possible to rewrite (\ref{ma4}) using the covariantised transverse derivative (\ref{sc3}). First, we express (\ref{new1}) in terms of this derivative as,
\be
\hat{\cal{L}}_{ab}=r_a \tilde{\nabla}_b - r_b\tilde{\nabla}_a
\ee
Taking its repeated product yields,
\ber
\frac{s}{2l^2}\hat{\cal{L}}_{ab}\hat{\cal{L}}^{ab}&=&\tilde{\nabla}_a\tilde{\nabla}^a+\frac{s}{l^2}r_a(\tilde{\nabla}_br^a)\tilde{\nabla}^b
\nonumber
\\
\nonumber
\\&=&\tilde{\nabla}_a\tilde{\nabla}^a+\frac{s}{l^2}r_a(\nabla_br^a)\nabla^b
\nonumber
\\
\nonumber
\\&=&\tilde{\nabla}_a\tilde{\nabla}^a=\tilde{\nabla}^2
\eer
where use has been made of (\ref{sc5}). Hence the following form of (\ref{ma4}) is obtained,
\be
(sl^2 \tilde{\nabla}^2 -2)\hat{A}_a=0
\ee
which can be interpreted as the non-abelian extension of (\ref{sc1}), subject to subsidiary and Lorentz gauge conditions.

     We conclude this section by discussing two issues; the possibilities of hyperbolic A(dS) space as infrared regulators and secondly, the A(dS)/CFT correspondence in our approach.

     It was shown in \cite{cw} that a space of constant negative curvature like the AdS space provided a good infrared regulator for euclidean quantum field theory. The method used there was also based on stereographic projection. Hence it is simple to relate the findings of \cite{cw} with our analysis. In fact the AdS gauge field equation of motion, which is the starting point in \cite{cw}, just corresponds to the $s=1$ version of (\ref{104.1}). There was an arbitrariness in the general solution of the Green function which was appropriately tuned to improve the asymptotic behaviour. Since the de-Sitter case implies $s=-1$ in (\ref{104.1}), these arguments go through and one expects that the infrared properties also improve here. This is presumably related to the fact that time slices of the de-Sitter space can be chosen to make the spatial volume finite thereby strongly affecting the infrared behaviour. Indeed it is precisely this property of the boundedness of volume that makes euclidean QED on the sphere manifestly infrared-finite \cite{adler}.

    Recently there have been several studies \cite{as,ljl} to extend the AdS/CFT correspondence \cite{mal} to include the de-Sitter space. While it is unclear whether the hypothesised dS/CFT correspondence is on the same footing as the AdS/CFT one, it is possible to show their equvalence at the algebric level. The point is that the isometry group $SO(2,n)$ (for AdS$_{n+1}$) or $SO(1,n+1)$ (for dS$_{n+1}$) acts on the boundary as the conformal group acting on Minkowski (euclidean) space. The same geometry as defined by (\ref{ds1}) is used. Light cone coordinates are then introduced to define the `projective boundary' of A(dS) space. The purported results  follow by considering the action of the group- say $SO(1,n+1)$- on the boundary points. We refer to \cite{jlp} for  relevant technical details.

\section {Stereographic Projections of the Dirac Equation on A(dS) space}

\bigskip

In order to discuss the stereographic projection of the Dirac equation it is first necessary to introduce the various Dirac matrices. In
the ordinary $D=4$  Minkowski space, these matices satisfy,
\be
\lbrace{\gamma_\mu , \gamma_\nu}\rbrace = \gamma_\mu \gamma_\nu + \gamma_\nu \gamma_\mu = 2\eta{_\mu}{_\nu}\,\,\,\,\,\,;\,\,\,\,\,\, \lbrace{\gamma_\mu,\gamma_4}\rbrace  =0 
\label{new5}
\ee
\be
\gamma_\mu{^\dagger} = \gamma_0 \gamma_\mu \gamma_0\,\,\,\,\,;\,\,\,\gamma_4{^\dagger}=\gamma_4 \,\,\,\,\,;\,\,\gamma_4=i\gamma_0\gamma_1\gamma_2\gamma_3  \,\,\,\,;\,\,\,(\gamma_4)^2=1
\label{new6}
\ee
Now we define the gamma matrices on the de Sitter (dS) space as,
\be
\Gamma_\mu = \gamma_\mu \gamma_4
\label{laltu3}
\ee
\be
\Gamma_4 = \gamma_4
\label{laltu4}
\ee
while, for the anti-de Sitter (AdS) space as,
\be
\Gamma_\mu = i\gamma_\mu \gamma_4
\label{laltu5}
\ee
\be
\Gamma_4 = \gamma_4
\label{laltu6}
\ee
These gamma matrices are defined so that, in either case, they obey the following properties,
\be
\lbrace{\Gamma_a , \Gamma_b }\rbrace = 2 s\eta{_a}{_b}
\label{laltu7}
\ee
and
\be
(r.\Gamma)^2 = l^2
\label{laltu8}
\ee
In order that the current satisfies a transversality condition like (\ref{s6}), we take its form on A(dS) space as, 
\be
\hat{j}_a = p  \frac{1}{2l} \bar{\hat{\Psi}} [ \Gamma . r , \Gamma_a ] \hat{\Psi}
\label{laltu9}
\ee
so that $r^a \hat{j}_a = 0$. Here, $p=1$ for dS space and $p=-i$ for AdS space.
The field, $ \hat{\Psi}$ is the Dirac spinor on the A(dS) space which is mapped to the Dirac spinor $\Psi$ on the Minkowski space through the following relation,
\be
\hat{\Psi} =  (1-\frac{x^2}{4l^2}) (1-\frac{x^\mu}{2l}\gamma_\mu)\Psi
\label{laltu10}
\ee
for dS space.  For AdS space the corresponding map is,
\be
\hat{\Psi} =  (1+\frac{x^2}{4l^2}) (1-i\frac{x^\mu}{2l}\gamma_\mu)\Psi
\label{laltu11}
\ee
From these relations, we define the adjoint spinor as,
\be
\bar{\hat{\Psi}} = \hat{\Psi}^\dagger \Gamma_0 =  (1-\frac{x^2}{4l^2})\bar{\Psi}(1+\frac{x^\mu}{2l}\gamma_\mu)
\label{laltu12}
\ee
for dS space and for AdS space as,
\be
\bar{\hat{\Psi}} = \hat{\Psi}^\dagger \Gamma_0 \Gamma_4 =   (1+\frac{x^2}{4l^2})\bar{\Psi}(1+i\frac{x^\mu}{2l}\gamma_\mu)
\label{laltu13}
\ee
This completes the stereographic mapping of the Dirac spinor. We now use these expressions to prove,
\be
\hat{j}_a = (1+s\frac{x^2}{4l^2})^2 K{_a}{^\mu}j_\mu
\label{new9}
\ee
which is the exact analog of (\ref{k}), apart from the conformal factor.

      Here the explicit calculation for dS space is given. Calculations for AdS space are similar.
For $a=\mu$ we have from (\ref{laltu9}),
\ber
\hat{j}_\mu &=& \frac{1}{2l}(1-\frac{x^2}{4l^2})^2 (r^\alpha [\gamma_\alpha , \gamma_\mu] - \frac{r^\alpha x^\lambda}{2l} [\gamma_\alpha , \gamma_\mu] \gamma_\lambda + 2 r^4 \gamma_\mu - \frac{2r^4x^\lambda}{2l}\gamma_\mu \gamma_\lambda +\frac{r^\alpha x^\nu}{2l}\gamma_\nu[\gamma_\alpha , \gamma_\mu] 
\nonumber
\\
\nonumber
\\&-&\frac{r^\alpha x^\lambda x^\nu}{4l^2}\gamma_\nu[\gamma_\alpha , \gamma_\mu]\gamma_\lambda                                                                                + \frac{2r^4x^\nu}{2l}\gamma_\nu \gamma_\mu - \frac{2r^4x^\lambda x^\nu}{4l^2}\gamma_\nu \gamma_\mu \gamma_\lambda)
\eer
Using the identity,
\be
\gamma_\mu \gamma_\nu \gamma_\alpha = \eta{_\mu}{_\nu}\gamma_\alpha - \eta{_\mu}{_\alpha}\gamma_\nu + \eta{_\nu}{_\alpha}\gamma_\mu - i\epsilon{_\mu}{_\nu}{_\alpha}{_\lambda}\gamma^\lambda \gamma_4
\label{i1}
\ee
 and the properties of gamma matrices ((\ref{new5}),(\ref{new6})) we get, 
\be
\hat{j}_\mu = (1-\frac{x^2}{4l^2})^2 K{_\mu}{^\nu}j_\nu
\ee
Similarly for $a=4$ one can show,
\be
\hat{j}_4 =  (1-\frac{x^2}{4l^2})^2 K{_4}{^\mu}j_\mu
\ee
These two can be written as,
\be
\hat{j}_a = (1-\frac{x^2}{4l^2})^2 K{_a}{^\mu}j_\mu
\ee
For AdS space this will be,
\be
\hat{j}_a = (1+\frac{x^2}{4l^2})^2 K{_a}{^\mu}j_\mu
\ee
This proves equation (\ref{new9}).
The implication of the conformal factor in (\ref{new9}) becomes evident when we introduce coupling with the gauge fields. This factor is necessary to ensure the form invariance of the interaction term in the action,
\be
\int d\Omega (\hat j_a \hat A^a) = \int d^4x (j_\mu A^\mu)
\label{m1}
\ee
This is verified by an explicit use of the various maps.
Indeed in D-dimensions, the map (\ref{new9}) reads as,
\be
\hat{j}_a = (1+s\frac{x^2}{4l^2})^{D-2} K{_a}{^\mu}j_\mu.
\label{133}
\ee
so that form invariance of the interaction term in the action as illustrated in (\ref{m1}) is valid in any dimensions.
The inverse map, computed from (\ref{new9}) by contracting with the Killing vector and using the identity (\ref{k1}), is given by,
\ber
j_\mu &=& (1+s\frac{x^2}{4l^2})^{-4} K_{a\mu} \hat{j}^a 
\nonumber
\\
\nonumber
\\&=& (1+s\frac{x^2}{4l^2})^{-2} \frac{\partial{r_a}}{\partial{x^\mu}} \hat{j}^a
\eer

           We have seen that the vector fields are mapped by the conformal Killing vectors. So, we can expect that the Dirac spinors are mapped by the conformal Killing spinors. To show this we define a transformation matrix $W$ such that,
\be
W =  (1+\frac{x^\mu}{2l}\gamma_\mu) 
\label{laltu14}
\ee
for dS space
and for AdS space,
\be
W =  (1+i\frac{x^\mu}{2l}\gamma_\mu)
\label{laltu15}
\ee
The adjoint of `$W$', following (\ref{laltu12}), is given by,
\be
\bar{W} = W^\dagger \Gamma_0 =  \Gamma_0 (1-\frac{x^\mu}{2l}\gamma_\mu)
\label{laltu16}
\ee
for dS space. For AdS space the corresponding definition follows from (\ref{laltu13}),
\be
\bar{W} = W^\dagger \Gamma_0 \Gamma_4 = \Gamma_0 \Gamma_4 (1-i\frac{x^\mu}{2l}\gamma_\mu)
\label{laltu17}
\ee
Therefore, for dS space we have,
\be
\bar{W}W = \Gamma_0 (1-\frac{x^2}{4l^2})
\label{bibhas15}
\ee
and the corresponding relation for AdS space is,
\be
\bar{W}W = \Gamma_0\Gamma_4 (1+\frac{x^2}{4l^2})
\label{bibhas16}
\ee
Using the above equations, the fermion maps (\ref{laltu10}) and (\ref{laltu11}) are written collectively as ,
\be
\Psi = (1+s\frac{x^2}{4l^2})^{-2} W \hat{\Psi}
\label{laltu18}
\ee
The maps for the adjoint spinors, on the other hand, are different. For dS space (\ref{laltu12}) becomes,
\be
\bar{\Psi} = -(1-\frac{x^2}{4l^2})^{-2}\hat{\bar{\Psi}} \Gamma_0 \bar{W}
\label{laltu19}
\ee
while for AdS space, (\ref{laltu13}) simplifies to,
\be
\bar{\Psi} =  (1+\frac{x^2}{4l^2})^{-2}\hat{\bar{\Psi}} \Gamma_0 \Gamma_4 \bar{W}
\label{laltu20}
\ee
The mapping (\ref{laltu18}) of fermions from A(dS) space to flat space has now been expressed in terms of `W' which is the conformal Killing spinor satisfying the equation,
\be
\partial_\mu W = \frac{1}{4} \gamma_\mu(\gamma^\sigma \partial_\sigma W)
\ee
This is analogous to (\ref{s10})  that defines the conformal Killing vector. 
The above relation yields a general definition for a conformal Killing spinor. Similar relations have appeared previously in the literature \cite{blau}.

         It is known that \cite{banerjee} there is a bilinear map connecting the conformal Killing vectors with the conformal Killing spinors. To see this  we have to write  (\ref{laltu9}) in a different form which is given below. 
 Using (\ref{laltu7}) and (\ref{laltu8}) one can show that 
\be
[\Gamma.r , \Gamma_a] = -2 (\eta{_a}{^b} - s \frac{r_a r^b}{l^2})\Gamma_b (\Gamma.r)
\label{laltu22}
\ee
Therefore the current (\ref{laltu9}) on A(dS) space is also expressed as,
\ber
&&\hat{j}_a = -p\frac{1}{l} \bar{\hat{\Psi}}(\eta{_a}{^b}-s\frac{r_a r^b}{l^2})\Gamma_b(\Gamma.r)\hat{\Psi}
\nonumber
\\
&&=\frac{-p}{l}\bar{\hat{\Psi}}\theta{_a}{^b}\Gamma_b\Gamma.r\hat{\Psi}
\label{laltu23}
\eer
Now with the help of (\ref{k2}), the above equation is written as,
\be
\hat{j}_a = -p\frac{1}{l}(1+s\frac{x^2}{4l^2})^{-2} \bar{\hat{\Psi}} K{_a}{_\mu}K{^b}{^\mu}\Gamma_b(\Gamma.r)\hat{\Psi}
\ee
Therefore, by (\ref{new9}) we have,
\be
 -p\frac{1}{l}(1+s\frac{x^2}{4l^2})^{-4} \bar{\hat{\Psi}} K{_a}{_\mu}K{^b}{^\mu}\Gamma_b(\Gamma.r)\hat{\Psi} = \bar{\Psi}K{_a}{^\mu} \gamma_\mu \Psi
\ee
Using the fermionic spinor maps (\ref{laltu10}) and (\ref{laltu12}) (for dS space), the above can be written as,
\be
\frac{1}{l}(1-\frac{x^2}{4l^2})^{-2} W K_{b\mu}\Gamma^b(\Gamma.r)\Gamma_0 \bar{W} = \gamma_\mu
\ee
Multiplying `$\bar{W}$' from left and then `$W$' from right on both sides of the above  equation and using (\ref{bibhas15}) we get,
\be
\bar{W}\gamma_\mu W = \frac{1}{l} K{_b}{_\mu}(\Gamma_0 \Gamma^b)(\Gamma.r)
\ee
for dS space. For AdS space, the calculation is similar and yields,
\be
\bar{W}\gamma_\mu W = \frac{i}{l} K{_b}{_\mu}(\Gamma_0 \Gamma_4 \Gamma^b)(\Gamma.r).
\ee
These are the cherished bilinear relations between the Killing spinors and the Killing vectors.

         Let us next consider the definition of the axial current. The analogoue of $\gamma_4$(chirality operator on Minkowski space) is given here by $\frac{1}{l}\Gamma . r$ since it satisfies,
\be
(\frac{1}{l} \Gamma .r)^2 = 1
\label{152.1}
\ee
and the anticommutator,
\be
\lbrace\frac{\Gamma.r}{l} , [\Gamma.r , \Gamma_a]\rbrace = 0
\ee
The projection operators are given by,
\be
P_{\pm} = \frac{1\pm\frac{\Gamma.r}{l}}{2},\,\,\,\,\,;\,\,\, P_{\pm}^{2}=P_{\pm},\,\,\,\,;\,\,\,\,P_{+}P_{-}=0
\ee
Hence the axial current is defined as,
\be
\hat{j}{_a}{_5} = -p\frac{1}{2l^2} \bar{\hat{\Psi}} [\Gamma.r , \Gamma_a] (\Gamma.r) \hat{\Psi}
\ee
This also satisfies the transversality condition $r^a \hat{j}_{a5} = 0$.
As was shown for the vector current, one can prove,
\be
\hat{j}{_a}{_5} 
= (1+s\frac{x^2}{4l^2})^2 K{_a}{^\mu} j{_\mu}{_5}
\label{laltu21}
\ee
where $j{_\mu}{_5} = \hat{\Psi} \gamma_\mu \gamma_4 \Psi$ is the axial current on the flat space.
This can also be wtitten (similar to \ref{laltu23}) as,
\be
\hat{j}{_a}{_5} = -p \bar{\hat{\Psi}}(\eta{_a}{^b}-s\frac{r_a r^b}{l^2})\Gamma_b\hat{\Psi}=-p\bar{\hat{\Psi}}\theta{_a}{^b}\Gamma_b\hat{\Psi}
\label{laltu24}
\ee

             Now we will write the Dirac operator and the Dirac equation on A(dS) space. For this we define,
\be
S{_a}{_b} =\frac{1}{4} [\Gamma_a , \Gamma_b]
\ee
Using the definition for $S_{ab}$ given above, one can show the algebric relation,
\be
S_{ab}L^{ab} = -\frac{1}{4}[\gamma_\mu,\gamma_\nu]L^{\mu\nu} + \gamma_\mu L^{4\mu}
\label{bibhas17}
\ee
for dS space. The corresponding relation for AdS space is,
\be
S_{ab}L^{ab} = \frac{1}{4}[\gamma_\mu,\gamma_\nu]L^{\mu\nu} - i\gamma_\mu L^{4\mu}
\ee
These relations are collectively expressed as,
\be
S_{ab}L^{ab}=\frac{s}{4}[\gamma_\mu,\gamma_\nu]L^{\mu\nu}+p\gamma_\mu L^{4\mu}
\ee
Again using  (\ref{s18a}) , (\ref{s18b}) and (\ref{v8}) we get the form of the angular momentum operator on the flat space as, 
\be
L{_\mu}{_\nu} = x_\mu \partial_\nu - x_\nu \partial_\mu
\label{162.1}
\ee
\be
L{_4}{_\mu} = -sl(1-s\frac{x^2}{4l^2}) \partial_\mu - \frac{1}{2l} x_\mu x^\nu \partial_\nu 
\label{163.1}
\ee
 Using the fermion maps (\ref{laltu10}), (\ref{laltu12}) and (\ref{bibhas17}) together with the expression for the angular momentum operator given above we obtain,
\be
\bar{\hat{\Psi}}(S_{ab}L^{ab}+2)\hat{\Psi} = l(1-\frac{x^2}{4l^2})^4 \bar{\Psi}\gamma^\mu \partial_\mu \Psi
\ee 
for dS space. The corresponding calculations for AdS space are exactly the same. For this one can show,
\be
\bar{\hat{\Psi}}(-S_{ab}L^{ab}+2)\hat{\Psi} = l(1+\frac{x^2}{4l^2})^4 \bar{\Psi}\gamma^\mu \partial_\mu \Psi
\ee 
So, in general we can write,
\be
\bar{\hat{\Psi}} (-s S{_a}{_b}L{^a}{^b}+2) \hat{\Psi} = l(1+s\frac{x^2}{4l^2})^4 \bar{\Psi}\gamma^\mu \partial_\mu \Psi
\label{new8}
\ee
This enables us to  convert the Dirac operator  into the psudospherical operator by the map,
\be
l\gamma^\mu \partial_\mu \longrightarrow -sS{_a}{_b}L{^a}{^b} + 2
\label{i2}
\ee

          The Dirac action on the flat Minkowski space is given by,
\be
S = -i \int d^4x \bar{\Psi} \gamma^\mu \partial_\mu \Psi
\ee
Using the map (\ref{conformal}) for the measure and (\ref{new8}), we obtain the following action for A(dS) space by a projection of the above flat action,
\be
S =  -\frac{i}{l} \int d\Omega \bar{\hat{\Psi}} (-sS{_a}{_b}L{^a}{^b} + 2)\hat{\Psi}
\label{A1}
\ee

      It is possible to show that the above action, exactly in analogy to the case of the gauge field, can be obtained from the standard action defined on the curved space which is taken as,
\be
S=\int d^4x (\surd{-g})\bar{\Psi}_c e^{\mu\nu}\gamma_\nu \nabla_\mu \Psi_c
\label{fer1}
\ee 
where $`\Psi_c$' is the Dirac spinor on the curved space and is connected to the flat space Dirac spinor through the conformal factor $(1+s\frac{x^2}{4l^2})^\frac{3}{2}$, which is required for the above action to be conformally invariant. The vielbein $e_{\mu\nu}$ is related to the flat Minkowski metric by the following relation,
\be
e_{\mu\nu}=(1+s\frac{x^2}{4l^2})^{-1}\eta_{\mu\nu}
\label{fer2}
\ee 
The covariant derivative $\nabla_\mu$ for a spinor is,
\be
\nabla_\mu=\partial_\mu+\frac{1}{2}\omega_{\mu\alpha\beta}\sigma^{\alpha\beta}
\label{fer3}
\ee
where the spin connection is defined as,
\be
\omega_{\mu\alpha\beta}=\frac{1}{2}(C_{\mu,\alpha\beta}-C_{\alpha,\beta\mu}-C_{\beta,\mu\alpha})
\label{fer4}
\ee
\be
C{^\nu}{_,}{_\alpha}{_\beta} = (\partial_\alpha e{_\beta}{^\nu}-\partial_\beta e{_\alpha}{^\nu})
\label{fer5}
\ee
and
\be
\sigma^{\alpha\beta}=\frac{1}{4}[\gamma^\alpha,\gamma^\beta]
\label{fer6}
\ee
Now using (\ref{conformal}), (\ref{new3}), (\ref{fer2}) and the transformation relation $\Psi_c=(1+s\frac{x^2}{4l^2})^{\frac{3}{2}}\Psi$, the first part of the action (\ref{fer1}) involving the ordinary derivative can be written as,
\ber
S_1 &=& -i\int d\Omega (1+s\frac{x^2}{4l^2})^4 (1+s\frac{x^2}{4l^2})^{-4} (1+s\frac{x^2}{4l^2})^{\frac{3}{2}}\bar{\Psi}(1+s\frac{x^2}{4l^2})
\nonumber
\\
\nonumber
& &\eta^{\mu\nu}
\gamma_\nu \partial_\mu[(1+s\frac{x^2}{4l^2})^{\frac{3}{2}}\Psi]
\\
\nonumber
\\
&=&-i\int d\Omega (1+s\frac{x^2}{4l^2})^4 \bar{\Psi}\gamma^\mu \partial_\mu\Psi-\frac{3is}{4l^2}\int d\Omega (1+s\frac{x^2}{4l^2})^3 \bar{\Psi}\gamma^\mu x_\mu \Psi
\label{fer7}
\eer
For the second part we have to first calculate the form of the spin connection.  Using (\ref{fer2}) and (\ref{fer5}) we have,
\ber
C_{\mu,\alpha\beta}&=&e_{\mu\nu} C{^\nu}{_,}{_\alpha}{_\beta}
\nonumber
\\
\nonumber
\\
&=&e_{\mu\nu}[\partial_\alpha(g^{\nu\lambda}e_{\beta\lambda})-\partial_\beta(g^{\nu\lambda}e_{\alpha\lambda})]
\nonumber
\\
\nonumber
\\
&=& \frac{s}{2l^2}(1+s\frac{x^2}{4l^2})^{-1}(x_\alpha\eta_{\beta\mu}-x_\beta\eta_{\alpha\mu})
\label{fer8}
\eer
Therefore, using the definition for the spin connection (\ref{fer4}), we obtain,
\be
\omega_{\mu\alpha\beta}= \frac{s}{2l^2}(1+s\frac{x^2}{4l^2})^{-1}(x_\alpha\eta_{\beta\mu}-x_\beta\eta_{\alpha\mu})
\label{fer9}
\ee
Putting all the results in the second part of the action we get,
\be
S_2=-\frac{is}{16l^2}\int d\Omega (1+s\frac{x^2}{4l^2})^3 \bar{\Psi}(x_\alpha\eta_{\beta\mu}-x_\beta\eta_{\alpha\mu})\gamma^\mu[\gamma^\alpha,\gamma^\beta]\Psi
\label{fer10}
\ee
Now, using the previous identity (\ref{i1}), we find,
\be
(x_\alpha\eta_{\beta\mu}-x_\beta\eta_{\alpha\mu})\gamma^\mu[\gamma^\alpha,\gamma^\beta]=-12x_\mu\gamma^\mu
\label{fer11}
\ee
Putting this result in (\ref{fer10}) we obtain,
\be
S_2 = \frac{3is}{4l^2}\int d\Omega (1+s\frac{x^2}{4l^2})^3 \bar{\Psi}\gamma^\mu x_\mu \Psi
\label{fer12}
\ee
Therefore, adding these two parts (\ref{fer7}) and (\ref{fer12}) of the action (\ref{fer1}) we get,
\be
S=S_1+S_2=-i\int d\Omega(1+s\frac{x^2}{4l^2})^4\bar{\Psi}\gamma^\mu \partial_\mu\Psi
\label{fer13}
\ee
Finally using (\ref{new8}) we obtain the cherished form,
\be
S=-\frac{i}{l}\int d\Omega \bar{\hat{\Psi}}(-sS_{ab}L^{ab}+2)\hat{\Psi}
\label{fer14}
\ee
which reproduces (\ref{A1}).

So the Dirac equation on A(dS) space is given by,
\be
(-s S{_a}{_b} L{^a}{^b} + 2)\hat{\Psi} = 0
\label{d}
\ee
For the hypersphere an analogous relation is given in \cite{adler,jackiw}.

For general D-dimensions  the Dirac equation on A(dS) space can be written as:
\be
 (-s S{_a}{_b} L{^a}{^b} + \frac{D}{2})\hat{\Psi} = 0
\label{171.1}
\ee
For $D=4$ we get back equation(\ref{d}).

     In presence of fermionic matter the flat space Yang-Mills action has the form,
\be
S=\int d^4x[-\frac{1}{4}Tr.(F_{\mu\nu}F^{\mu\nu})-i\bar{\Psi}\gamma^\mu(\partial_\mu-ieA_\mu)\Psi]
\ee
Using appropriate expressions for each piece, the stereographically projected action on the A(dS) space becomes,
\be
\hat{S}=\int d\Omega[-\frac{s}{12l^2}Tr.(\hat{F}_{abc}\hat{F}^{abc})-\frac{i}{l}\hat{\bar{\Psi}}(-sS_{ab}L^{ab}+2)\hat{\Psi}-e\hat{A}_a\hat{j}^a]
\ee
where, $\hat{j}^a$ is defined in (\ref{laltu9}) and we have used (\ref{new9}) to project the interaction term. Hence, in the presence of gauge fields, equation (\ref{d}) becomes,
\be
[\frac{is}{l} (S_{ab})_{ij}\eta{_n}{^m}L^{ab}-\frac{2i}{l}\eta{_n}{^m}\delta_{ij}+\frac{2ep}{l}(S_{ab})_{ij}r^b(\hat{A^{\lambda}}^a)(T^\lambda){_n}{^m}]\hat{\Psi}_{jm}=0
\ee

      By looking at the Dirac equation (\ref{d} or \ref{171.1}) one might be tempted to interpret the numerical factor as a mass term on A(dS) space. But this is not true. It is seen from (\ref{i2}) that this Dirac operator was obtained from a projection of the massless Dirac operator on the flat space. Also, it is equivalent to the massless Dirac operator (\ref{fer1}) defined on a curved background.

     Yet another way is to use chiral invariance. In the flat space the mass term breaks this invariance in the sence that the anticommuting property  $\{\gamma^\mu\partial_\mu,\gamma_5\} = 0$ of the massless Dirac operator no longer holds. In A(dS) space the analogue of $\gamma_5$ is $\frac{\Gamma.r}{l}$, as already discussed. We now compute the anticommutator of the Dirac operator appearing in (\ref{171.1}) with $\frac{\Gamma.r}{l}$ and show that it vanishes, thereby reconferming that the numerical factor should not be interpreted as a mass term. The de-Sitter space calculation is given below. The calculation for AdS space is exactly similar. Now, 
\be
\{S.L+\frac{D}{2},\frac{\Gamma.r}{l}\}\phi=\frac{1}{l}\{S.L,\Gamma.r\}\phi+D\frac{\Gamma.r}{l}\phi
\label{1.1}
\ee
The first term of the above equation yields,
\ber
\frac{1}{l}\{S.L (\Gamma.r\phi)+\Gamma.r S.L\phi\}
=\frac{1}{l}\{S_{ab}(L^{ab}\Gamma.r)\phi+S_{ab}\Gamma.rL^{ab}\phi+\Gamma.r S.L\phi\}
\label{1.2}
\eer
Now the last two terms of the above cancel each other, which is shown below.
\ber
&&S_{ab}\Gamma.r L^{ab}+\Gamma.r S.L
\nonumber
\\
&=&\frac{1}{4}[\Gamma_a,\Gamma_b]\Gamma.r L^{ab}+\frac{1}{4}\Gamma.r[\Gamma_a,\Gamma_b]L^{ab}
\nonumber
\\
&=&\frac{1}{2}\Gamma_a\Gamma_b\Gamma.r(r^a\partial^b-r^b\partial^a)+\frac{1}{2}\Gamma.r\Gamma_a\Gamma_b(r^a\partial^b-r^b\partial^a)
\nonumber
\\
&=&\frac{1}{2}[(\Gamma.r)\Gamma_a(\Gamma.r)\partial^a-\Gamma_a(\Gamma.r)^2\partial^a+(\Gamma.r)^2\Gamma_a\partial^a-(\Gamma.r)\Gamma_a(\Gamma.r)\partial^a]=0
\label{1.3}
\eer
where we have used (\ref{152.1}) in the last line.

  Now,
\ber
&&S_{ab}L^{ab}(\Gamma.r)
\nonumber
\\
&&=S_{\mu\nu}L^{\mu\nu}(\Gamma.r)+S_{D\mu}L^{D\mu}(\Gamma.r)
\nonumber
\\
&&=\frac{1}{2}[\gamma_\nu\gamma_\mu L^{\mu\nu}(\Gamma.r)-\gamma_\mu L^{D\mu}(\Gamma.r)]
\nonumber
\\
&&=\frac{1}{2}[\gamma_\nu\gamma_\mu\gamma_\alpha\gamma_D L^{\mu\nu}r^\alpha+\gamma_\nu\gamma_\mu\gamma_D L^{\mu\nu}r^D+\gamma_\mu\gamma_\nu\gamma_D L^{D\mu}r^\nu+\gamma_\mu\gamma_D L^{D\mu}r^D]
\label{1.4}
\eer
Using the expressions for $L^{\mu\nu}$(\ref{162.1}), $L^{D\mu}$(\ref{163.1}), $r^\mu$ and $r^D$(\ref{ds3}) we have,
\ber
&&L^{\mu\nu}r^\alpha=\frac{1}{1-\frac{x^2}{4l^2}}(x^\mu\eta^{\nu\alpha}-x^\nu\eta^{\mu\alpha})
\nonumber
\\
&&L^{\mu\nu}r^D=0
\nonumber
\\
&&L^{D\mu}r^\nu=-\frac{l(1+\frac{x^2}{4l^2})}{1-\frac{x^2}{4l^2}}\eta^{\mu\nu}
\nonumber
\\
&&L^{D\mu}r^D=\frac{x^\mu}{1-\frac{x^2}{4l^2}}
\label{1.5}
\eer
Therefore,
\ber
&&S_{ab}L^{ab}(\Gamma.r)
\nonumber
\\
&&=\frac{(1-D)x^\mu\gamma_\mu\gamma_D}{1-\frac{x^2}{4l^2}}+D\frac{l(1+\frac{x^2}{4l^2})}{1-\frac{x^2}{4l^2}}\gamma_D-\frac{x^\mu}{1-\frac{x^2}{4l^2}}\gamma_\mu\gamma_D
\nonumber
\\
&&=(1-D)r^\mu\Gamma_\mu-Dr^D\Gamma_D-r^\mu\Gamma_\mu
\nonumber
\\
&&=-D(\Gamma.r)
\label{1.6}
\eer
Using all these results, we obtain,
\be
\{S.L+\frac{D}{2},\frac{\Gamma.r}{l}\}=0
\label{1.7}
\ee
For AdS space this will be $\{-S.L+\frac{D}{2},\frac{\Gamma.r}{l}\}=0$. So, in general we obtain,
\be
\{-sS.L+\frac{D}{2},\frac{\Gamma.r}{l}\}=0
\label{1.8}
\ee
This shows that $\frac{D}{2}$ cannot be the mass term and the free Dirac operator for massless fermion on general D-dimensional A(dS) space is $\frac{1}{l}(-sS.L+\frac{D}{2})$.

    Now if we include a mass term, then the massive Dirac action on A(dS) space for 4-dimensions can be written as,
\be
S=\int d\Omega \bar{\hat{\Psi}}[-\frac{i}{l}(-sS.L+2)+\nu]\hat{\Psi}
\label{1.9}
\ee
where `$\nu$' plays the role of mass of fermion on A(dS) space. `$\nu$' can be determined by our projection method. The mass term of the flat space action is given by $m\bar{\Psi}\Psi$. Using the projection for the spinor field (\ref{laltu18}), (\ref{laltu19}) and (\ref{laltu20}) we obtain,
\be
m\bar{\Psi}\Psi=m(1+s\frac{x^2}{4l^2})^{-3}\bar{\hat{\Psi}}\hat{\Psi}
\label{1.10}
\ee
So the Dirac action on A(dS) space for the mass sector is,
\ber
S_m&=&\int d\Omega(1+s\frac{x^2}{4l^2})^4 .m (1+s\frac{x^2}{4l^2})^{-3}\bar{\hat{\Psi}}\hat{\Psi}
\nonumber
\\
&=&\int d\Omega m(1+s\frac{x^2}{4l^2}) \bar{\hat{\Psi}}\hat{\Psi}
\label{1.11}
\eer
Therefore the mass of the fermion on A(dS) space is given by,
\be
\nu=m(1+s\frac{x^2}{4l^2})
\label{1.12}
\ee
In the flat space limit $l\rightarrow\infty$ we get back the usual mass.

     It may be mentioned that so far the analysis has been basically classical. Extension to quantum field theory is quite nontrivial. In the next section we will take up this issue in some detail where an explicit calculation of the quantum anomaly is presented and its equivalence is established with our stereographic projection approach.

     There is another point to consider. Any quantum effect on the A(dS) space need not have a counterpart on the flat space. For instance, a quantum field theory on dS space will reveal the analogue of Hawking radiation because of the cosmological horizon. However, it appears highly unlikely that this physical effect would naturally connect to a corresponding feature of the projected theory on flat space.

    As a simple illustration consider the factorisation of the wave operator for Bose particles in terms of the fermionic wave operator that holds in flat space,
\be
(i\gamma.\p)(i\gamma.\partial)=-\Box
\ee
This dose not have an analogue on the A(dS) space. As we have shown, here the Bose operator (in four dimensions) is $(L_{ab}L^{ab}-4)$ while the Fermi operator is $(-sS_{ab}L^{ab}+2)$. It is now possible to show \cite{cw, adler},
\be
(-sS_{ab}L^{ab}+2)(-sS_{ab}L^{ab}+1)=-\frac{1}{2}(L_{ab}L^{ab}-4)
\ee
The simple factorisation in flat space therefore dose not hold in the A(dS) space.

\section {Chiral  Anomalies on A(dS) Space}

\bigskip

In this section we discuss the structure of chiral anomalies on A(dS) space. First, an explicit evaluation of the axial U(1) anomaly is computed in the path integral approach. This result is next reproduced by an appropriate stereographic map of the usual Adler-Bell-Jackiw \cite{donnell1} flat space expression. This shows that the methods developed here are meaningful for considering quantum effects. Finally, we obtain the non-abelian chiral anomalies by our projection technique.

\subsection {The Axial Anomaly in the Path Integral Formalism}

          In a series of papers, Fujikawa \cite{kawa} has shown how the chiral anomalies encountered in perturbation theory may be derived in a path integral framework. In this approach the anomalous behaviour of Ward-Takahashi identites is traced to the Jacobian factor arising from the noninvariance of the path integral measure under chiral transformations. Here we compute the chiral anomaly in the A(dS) space using this method. We follow the approach of \cite{donnell2} where the analysis has been done on the sphere. In our case an analytic continuation of the A(dS) pseudosphere is implied. The advantage of working on the compact space is that it admits a complete set of familiar basis functions, namely the spherical harmonics. The generating functional is,
\be
Z(\eta,\bar{\eta},\chi_a)=\int d\mu exp(\int d\Omega[{\cal{L}}+\bar{\eta}\hat{\Psi}+\bar{\hat{\Psi}}\eta+\chi.\hat{A}])
\label{2.1}
\ee      
where $d\mu = [d\bar{\hat{\Psi}}][d\hat{\Psi}][d\hat{A}_a]$ is the functional measure including the Faddeev-Popov factor and $d\Omega$ is the volume element.

  Now the action (in four dimensions) is given by,
\ber
&&S=\int d\Omega [\bar{\hat{\Psi}}\frac{1}{l}(-sS.L+2)\hat{\Psi}+ie\hat{A}_a\hat{j}^a]
\nonumber
\\
&&=\int d\Omega[\bar{\hat{\Psi}}\frac{1}{l}(-sS.L+2)\hat{\Psi}-iep\hat{A}_a\bar{\hat{\Psi}}\theta^{ab}\Gamma_b\frac{\Gamma.r}{l}\hat{\Psi}]
\nonumber
\\
&&=\int d\Omega \bar{\hat{\Psi}}[\frac{1}{l}(-sS.L+2)-ig\hat{A}_a\theta^{ab}\Gamma_b\frac{\Gamma.r}{l}]\hat{\Psi}
\nonumber
\\
&&=\int d\Omega \bar{\hat{\Psi}}[Q-ig\theta^{ab}\Gamma_b\frac{\Gamma.r}{l}\hat{A}_a]\hat{\Psi}
\label{2.2}
\eer
where in the second line use has been made of (\ref{laltu23}) and $Q=\frac{1}{l}(-sS.L+2)$, $g=ep$.

   Here we will show the calculation for dS space (i.e. s=-1). AdS space calculation is similar to it. Following Fujikawa, let $\phi_n$ be a complete set of eigenfunction for the Dirac operator
\be
D_A = Q-ig\theta^{ab}\Gamma_b\frac{\Gamma.r}{l}\hat{A}_a
\label{2.3}
\ee
i.e. 
\ber
&&D_A\phi_n=\lambda_n\phi_n
\nonumber
\\
&&\int d\Omega \phi{{^\dagger}{_n}}(r)\phi_m(r)=\delta_{nm}
\label{2.4}
\eer
Under the chirality transformation $\hat{\Psi}\rightarrow e^{i\e(r)\frac{\Gamma.r}{l}}\hat{\Psi}$, $\bar{\hat{\Psi}}\rightarrow\bar{\hat{\Psi}}e^{i\e(r)\frac{\Gamma.r}{l}}$ the functional measure transforms as $d\mu\rightarrow d\mu exp[-2i\int d\Omega \e(r){A}(r)]$, where
\be
A(r)=\sum_{n}\phi{{^\dagger}{_n}}\frac{\Gamma.r}{l}\phi_n
\label{2.5}
\ee
and the lagrangian transforms as ${\cal{L}}\rightarrow{\cal{L}}-i\e [L^{ab}(\bar{\hat{\Psi}}S_{ab}\frac{\Gamma.r}{l^2}\hat{\Psi})-2\bar{\hat{\Psi}}\frac{\Gamma.r}{l^2}\hat{\Psi}]$. Now the requirement of invariance of the generating functional under chiral transformation gives the correct anomaly equation, which turns out to be,
\be
L^{ab}[\bar{\hat{\Psi}}S_{ab}\frac{\Gamma.r}{l^2}\hat{\Psi}]-2\bar{\hat{\Psi}}\frac{\Gamma.r}{l^2}= -2A(r)
\ee
Multiplying the above by $r_c$ we obtain the final form of the anomaly equation as,
\be
\hat{j}_{c5}-L_{cb}\hat{j}^{b5}= 2r_cA(r)
\label{ano}
\ee
where $\hat{j}_{a5}$ is given by equation (\ref{laltu24}).
$A(r)$ is the anomaly factor which will now be explicitly computed.

 The conditionally convergent sum in $A(r)$ (\ref{2.5}) is evaluated by regularizing large eigenvalues and changing to the free spinor harmonic basis. The spinor harmonics $\Psi^{(\mu)}{_{l^{'}}}{_m}{_s}(r)$ are constructed to be orthonormalized eigenfunctions of the free massless Dirac operator $Q=\frac{1}{l}(S.L+2)$; i.e.
\ber
&&Q \Psi^{(\mu)}{_{l^{'}}}{_m}{_s}=\mu\Psi^{(\mu)}{_{l^{'}}}{_m}{_s},\,\,\,;\,\,\mu=-(l^{'}+2)(l^{'}+1),
\nonumber
\\
&&\Psi^{(\mu)}{_{l^{'}}}{_m}{_s}=P^{(\mu)}Y_{{l^{'}}m}\chi_s,
\nonumber
\\
&&P^{-({l^{'}}+2)}=\frac{{l^{'}}+3-S.L}{2l^{'}+3},
\nonumber
\\
&&P^{(l^{'}+1)}=\frac{1+S.L}{2l^{'}+3}
\label{2.6}
\eer
In the above set $Y_{l^{'}m}(r)$ are four dimensional spherical harmonics which satisfy,
\ber
&&\int d\Omega Y_{{l^{'}}_1 m_1}(r)Y_{{l^{'}}_2 m_2}(r)=\delta_{{l^{'}}_1{l^{'}}_2}\delta_{m_1m_2},
\nonumber
\\
&&\sum_{m} Y_{{l^{'}}m}(r)Y_{{l^{'}}m}(r^{'})= \frac{2l^{'}+3}{4\pi^{\frac{5}{2}}}\Gamma(\frac{3}{2})C{_{l^{'}}}{^{\frac{3}{2}}}(\frac{r.r^{'}}{l^2}),
\nonumber
\\
&&\sum_{l^{'}m} Y_{l^{'}m}(r)Y_{l^{'}m}(r^{'})=\delta(r-r^{'})
\label{2.7}
\eer
where the index `$m$' actually stands for the three `magnetic' quantum numbers and the `$\chi_s$' are constant orthonormal $2^{2}$ componant spinors. The $C{_{l^{'}}}{^{(\frac{3}{2})}}$ are Gegenbauer polynomials.

 Therefore,
\ber
A(r)&=&\lim_{M\rightarrow \infty}\sum_{n} \phi^{\dagger}{_n}(r)\frac{\Gamma.r}{l}e^{-(\frac{\lambda_n}{M})^2} \phi_n(r)
\nonumber
\\
&=&\lim_{M\rightarrow\infty}\sum_{{l^{'}}\mu ms}  \Psi^{(\mu)}{_{l^{'}}}{_m}{_s}{^\dagger}(r)\frac{\Gamma.r}{l} e^{-(\frac{D_A}{M})^2}\Psi^{(\mu)}{_{l{'}}}{_m}{_s}(r)
\label{2.8}
\eer
Now from (\ref{2.3}),
\ber
{D_A}^2 &=& Q^2 -ig(S.L+2)\theta^{ab}\Gamma_b\frac{\Gamma.r}{l^2}\hat{A}_a
\nonumber
\\
&-& ig\theta^{ab}\Gamma_b\frac{\Gamma.r}{l^2}\hat{A}_a(S.L+2)-g^2(\theta^{ab}\Gamma_b\frac{\Gamma.r}{l}\hat{A}_a)^2
\label{2.9}
\eer
In the previous section we have seen that $\{S.L+2,\frac{\Gamma.r}{l}\}=0$ (equation(\ref{1.7})). Also, one can show another anticommuting relation,
\be
\{\theta^{ab}\Gamma_b,\frac{\Gamma.r}{l}\}=0.
\label{2.10}
\ee
Using these anticommuting relations, (\ref{2.9}) can be written as,
\ber
{D_A}^2&=&Q^2-ig\frac{\Gamma.r}{l^2}(S.L+2)\theta^{ab}\Gamma_b\hat{A}_a
\nonumber
\\
&+&ig\frac{\Gamma.r}{l^2}\theta^{ab}\Gamma_b\hat{A}_a(S.L+2)+g^2(\theta^{ab}\Gamma_b\hat{A}_a)^2
\nonumber
\\
&=&Q^2-ig\frac{\Gamma.r}{l^2}S.L\theta^{ab}\Gamma_b\hat{A}_a+g^2(\theta^{ab}\Gamma_b\hat{A}_a)^2+ig\frac{\Gamma.r}{l^2}\theta^{ab}\Gamma_b\hat{A}_aS.L
\label{2.11}
\eer
Also, after some simplifications one can show,
\be
\Gamma.r\theta^{ab}\Gamma_b\hat{A}_aS.L=\hat{A}^ar_b\Gamma_a\Gamma_cL^{bc}
\ee
Therefore,
\ber
A(r)&=&\lim_{M\rightarrow\infty} \sum_{{l^{'}}\mu ms} \Psi^{(\mu)}{_{l^{'}}}{_m}{_s}(r)\frac{\Gamma.r}{l}exp\{-M^{-2}[Q^2-ig\frac{\Gamma.r}{l^2}S.L\theta^{ab}\Gamma_b\hat{A}_a
\nonumber
\\
&+&g^2(\theta^{ab}\Gamma_b\hat{A}_a)^2+ig\frac{\Gamma.r}{l^2}\theta^{ab}\Gamma_b\hat{A}_aS.L]\}\Psi^{(\mu)}{_{l^{'}}}{_m}{_s}(r)
\nonumber
\\
&=&\lim_{{M\rightarrow\infty},{ r\rightarrow r^{'}}} \sum_{l^{'}} Tr[\frac{\Gamma.r}{l}exp\{-M^{-2}(Q^2-ig\frac{\Gamma.r}{l^2}S.L\theta^{ab}\Gamma_b\hat{A}_a
\nonumber
\\
&+&g^2(\theta^{ab}\Gamma_b\hat{A}_a)^2+\frac{ig}{l^2}\hat{A}^ar_b\Gamma_a\Gamma_cL^{bc}\}]
\nonumber
\\
&&\times\frac{\Gamma(\frac{3}{2})}{4\pi^{\frac{5}{2}}}(2l^{'}+3)C{_{l^{'}}}{^{(\frac{3}{2})}}(\frac{r.r^{'}}{l^2})
\label{2.12}
\eer
Now one can check the following trace relation,
\be
Tr(\Gamma_a\Gamma_b\Gamma_c\Gamma_d\Gamma_e)=-4i \e_{abcde}
\label{2.13}
\ee
for four dimensions. The behaviour of the summand in (\ref{2.12}) for large $l^{'}$ is ${l^{'}}^{3}e^{-\frac{{l^{'}}^2}{M^2}}$. Together with the above trace relation and (\ref{2.10}) we can eliminate several terms in the exponentials.
Then (\ref{2.12}) simplifies to,
\ber
A(r)&=&\lim_{{M\rightarrow\infty},{ r\rightarrow r^{'}}}  \sum_{l^{'}} e^{-(\frac{l^{'}}{M})^2} C{_{l^{'}}}{^{(\frac{3}{2})}}(\frac{r.r^{'}}{l^2})\frac{\Gamma(\frac{3}{2})}{4\pi^{\frac{5}{2}}} (2l^{'}+3)
\nonumber
\\
&&\times Tr[\frac{\Gamma.r}{l}exp\{-\frac{1}{M^2}(-ig\frac{\Gamma.r}{l^2}S.L\theta^{ab}\Gamma_b\hat{A}_a+g^2(\theta^{ab}\Gamma_b\hat{A}_a)^2)\}]
\nonumber
\\
&=&\frac{\Gamma(\frac{3}{2})}{4\pi^{\frac{5}{2}}} \lim_{M\rightarrow\infty}\sum_{l^{'}} e^{-(\frac{{l^{'}}}{M})^2}(2l^{'}+3)\frac{\Gamma(l^{'}+3)}{\Gamma(l^{'}+1)\Gamma(3)}
\nonumber
\\
&&\times Tr[\frac{\Gamma.r}{l}exp\{-\frac{1}{M^2}(-ig\frac{\Gamma.r}{l^2}S.L\theta^{ab}\Gamma_b\hat{A}_a+g^2(\theta^{ab}\Gamma_b\hat{A}_a)^2)\}]
\label{2.14}
\eer
where we have used $C{_a}{^b}(1)={}^{(a+2b-1)}C{_{a}}$ in the last line. The sum in (\ref{2.14}) gives,
\ber
&&\frac{\Gamma{(\frac{3}{2})}}{2\pi^{\frac{5}{2}}\Gamma(3)}\sum_{l^{'}} e^{-(\frac{{l^{'}}}{M})^2} (l^{'}+\frac{3}{2})(l^{'}+2)(l^{'}+1)
\nonumber
\\
&=&\frac{2}{(4\pi)^2 }M \int dl^{'} e^{-{l^{'}}^2}[(Ml^{'})^{3}+O(Ml^{'})^{2}]
\nonumber
\\
&=&\frac{1}{(4\pi)^2 }M^{4}[1+O(\frac{1}{M})]
\label{2.15}
\eer
Expansion of the exponential in (\ref{2.14}) gives terms like,
\ber
&&\frac{1}{t!}(-\frac{1}{M^2})^t Tr[\frac{\Gamma.r}{l}\{-ig\frac{\Gamma.r}{l^2}S.L\theta^{ab}\Gamma_b\hat{A}_a+g^2(\theta^{ab}\Gamma_b\hat{A}_a)^2\}^t]
\nonumber
\\
&=&\frac{1}{t!}(-\frac{1}{M^2})^t Tr[\frac{\Gamma.r}{l}\{-ig\frac{\Gamma.r}{l^2}S.L\theta^{ab}\Gamma_b\hat{A}_a+g^2(\Gamma.\hat{A})^2\}^t]
\nonumber
\\
&=&\frac{1}{t!}(\frac{ig}{M^2})^t Tr[\frac{\Gamma.r}{l}\{\frac{\Gamma.r}{l^2}S.L\theta^{ab}\Gamma_b\hat{A}_a+ig(\Gamma.\hat{A})^2\}
\nonumber
\\
&&\{\frac{\Gamma.r}{l^2}S.L\theta^{ab}\Gamma_b\hat{A}_a+ig(\Gamma.\hat{A})^2\}^{t-1}]
\label{2.16}
\eer
Using the value of $\theta^{ab}$ and $S_{ab}$ the terms in the trace simplify to,
\be
\frac{\Gamma.r}{l}\{\frac{\Gamma.r}{l^2}S.L\theta^{ab}\Gamma_b\hat{A}_a+ig(\Gamma.\hat{A})^2\}=\frac{1}{2l}\Gamma_a\Gamma_b\Gamma_c(L^{ab}\hat{A}^c-2igr^a\hat{A}^b\hat{A}^c)
\label{2.17}
\ee
and
\be
\frac{\Gamma.r}{l^2}S.L\theta^{ab}\Gamma_b\hat{A}_a+ig(\Gamma.\hat{A})^2
=-\frac{r_d}{l^2}\Gamma_e\Gamma_f(L^{de}\hat{A}^f-igr^d\hat{A}^e\hat{A}^f)
\label{2.18}
\ee
where for the last relation we have used,
\ber
r^a\Gamma_a\Gamma_b\Gamma_c\Gamma_dL^{bc}\hat{A}^d
&=&r^a(-2\eta_{ab}-\Gamma_b\Gamma_a)\Gamma_c\Gamma_dL^{bc}\hat{A}^d
\nonumber
\\
&=&-2r_b\Gamma_c\Gamma_dL^{bc}\hat{A}^d
-\Gamma_b\Gamma.r\Gamma_c\Gamma_d(r^b\partial^c-r^c\partial^b)\hat{A}^d
\nonumber
\\
&=&-2r_b\Gamma_c\Gamma_dL^{bc}\hat{A}^d
\eer
since $(\Gamma.r)^2=l^2$.
So, R.H.S of (\ref{2.16}) becomes,
\ber
\frac{1}{t!}(-1)^{t-1}(\frac{ig}{M^2})^t Tr[\frac{1}{2l}\Gamma_a\Gamma_b\Gamma_c(L^{ab}\hat{A}^c-2igr^a\hat{A}^b\hat{A}^c)\{\frac{r_d}{l^2}\Gamma_e\Gamma_f(L^{de}\hat{A}^f-igr^d\hat{A}^e\hat{A}^f)
\}^{t-1}]
\label{2.19}
\eer
Now looking at (\ref{2.15}) and (\ref{2.19}) and considering the limit $M\rightarrow\infty$ it is seen that  only the $t=2$ term contrubutes in (\ref{2.19}). Therefore,
\ber
A(r)&=&\frac{1}{4(4\pi)^2}g^2 Tr[\Gamma_a\Gamma_b\Gamma_c(L^{ab}\hat{A}^c-2ig\frac{r^a}{l}\hat{A}^b\hat{A}^c)\frac{r_d}{l^3}\Gamma_e\Gamma_f(L^{de}\hat{A}^f-ig\frac{r^d}{l}\hat{A}^e\hat{A}^f)]
\nonumber
\\
&=&-i\frac{g^2}{16\pi^2l^3}\e_{abcef}[r_d(L^{ab}\hat{A}^c)(L^{de}\hat{A}^f)+igl(L^{ab}\hat{A}^c)\hat{A}^e\hat{A}^f
\nonumber
\\
&-&2ig\frac{r^a}{l}\hat{A}^b\hat{A}^cr_dL^{de}\hat{A}^f+2r^a\hat{A}^b\hat{A}^c\hat{A}^e\hat{A}^f]
\nonumber
\\
&=&-i\frac{g^2}{16\pi^2l^3}\e_{abcef}r_d(L^{ab}\hat{A}^c)(L^{de}\hat{A}^f)
\label{2.20}
\eer
where in the last line antisymmetricity of $\e_{abcef}$ has been used. 
Now from (\ref{v12}) we obtain (for abelian case),
\be
\hat{F}^{abc}r_d\hat{F}^{def} = 6\e_{abcef} L^{ab}\hat{A}^cr_dL^{de}\hat{A}^f+3\e_{abcef}L^{ab}\hat{A}^cr_dL^{ef}\hat{A}^d
\label{2.21}
\ee
Again, multiplying (\ref{v14}) from left by $\e_{abcef}$ and from right by $r_dL^{ef}\hat{A}^d$ we obtain,
\be
\e_{abcef}(L^{ab}K^{c\mu})r_dL^{ef}\hat{A}^d=0
\label{2.22}
\ee
Therefore the last term in (\ref{2.21}) reduces to,
\ber
&&\e_{abcef}(L^{ab}\hat{A}^c)r_d(L^{ef}\hat{A}^d)
\nonumber
\\
&=&\e_{abcef}(L^{ab}K^{c\nu}A_\nu)r_d(L^{ef}\hat{A}^d)
\nonumber
\\
&=&\e_{abcef}(L^{ab}K^{c\nu})A_\nu r_dL^{ef}\hat{A}^d+\e_{abcef}K^{c\nu}(L^{ab}A_\nu)r_dL^{ef}\hat{A}^d
\nonumber
\\
&=&\e_{abcef}K^{c\nu}(L^{ab}A_\nu)r_d(L^{ef}K^{d\mu})A_\mu+\e_{abcef}K^{c\nu}(L^{ab}A_\nu)r_dK^{d\mu}(L^{ef}A_\mu)=0
\label{2.23}
\eer
where we have used (\ref{2.22}) and $r_a K^{a\mu}=0$.
So (\ref{2.21}) yields,
\be
\hat{F}^{abc}r_d\hat{F}^{def} = 6\e_{abcef} L^{ab}\hat{A}^cr_dL^{de}\hat{A}^f
\ee
Substituting this in (\ref{2.20}) we obtain,
\ber
A(r)&=&-i\frac{g^2}{96\pi^2l^3}\e_{abcef} \hat{F}^{abc}r_d\hat{F}^{def} 
\nonumber
\\
&=&-i\frac{g^2}{96\pi^2l^3}r_a\e_{bcdef} \hat{F}^{abc}\hat{F}^{def} 
\eer
Putting everything in (\ref{ano}), we get the anomaly equation,
\be
\hat{j}_{g5}-L_{gb}\hat{j}^{b5}=-\frac{ig^2}{48\pi^2 l^3}r_gr_a\e_{bcdef}\hat{F}^{abc}\hat{F}^{def}.
\label{m12d}
\ee

           We now reproduce the above relation by stereographically projecting the familiar  Adler-Bell-Jackiw anomaly \cite{donnell1} on the A(dS) pseudosphere. For the axial current, employing a gauge invariant regularisation, the familiar result on the flat space
is known to be \cite{donnell1},
\be
\p_\mu j^{\mu 5} = {ig^2\over {16\pi^2}}\e_{\mu\nu\lambda\rho}F^{\mu\nu}F^{\lambda\rho}
\label{anomaly4}
\ee

Using (\ref{new}) and the definition of the current (\ref{new9}) (appropriately 
interpreted for the axial vector currents), it is possible to obtain the
identification,
\be
 r^a L_{ab} \hat j^{b5} = s l^2 \Big(1+s\frac{x^2}{4l^2}\Big)^4 \p_\mu j^{\mu 5}
\label{m9}
\ee 
In getting at the final result, use was made of the identity
(\ref{k3}). 
This provides a map for one side of (\ref{anomaly4}). To obtain an analogous form for the other side,
 it is necessary to consider the completely antisymmetric
tensor $\e_{\mu\nu\lambda\rho}$ whose value is the same in all systems. 

In order to provide a mapping among the $\epsilon$-tensors in the two spaces, we adopt the same rule (\ref{v11}) used for defining the antisymmetric field tensor.
However there is a slight subtlety. Strictly speaking, this Levi-Civita epsilon is a tensor density.
Hence its transformation law is modified by an appropriate conformal (weight) factor,
\be
\e_{abcde}=\frac{1}{l} \Big(1 + s\frac{x^2}{4l^2}\Big)^{-4}  
\Big(r_a K_b^\mu K_c^\nu K_d^\lambda K_e^\rho + cyclic\,\,\,\, permutations\,\,\,in\,\,\,(a, b, c, d, e) \Big)\e_{\mu\nu\lambda\rho}
\label{m5}
\ee
It is possible to verify the above relation by an explicit calculation, taking the convention
that both the epsilons are $+1 (-1)$ for any even (odd) permutation of distinct entries $(0,1, 2, 3, 4 )$ in that order.

     The inverse relation is obtained from (\ref{m5}) by appropriate contractions and exploiting the identity (\ref{k1}),
\ber
\e_{\mu\nu\lambda\rho}&=& s\frac{1}{l}(1+s\frac{x^2}{4l^2})^{-4} r_a K_{b\mu} K_{c\nu} K_{d\lambda} K_{e\rho} \e^{abcde}
\nonumber
\\
\nonumber
\\&=&\frac{s}{l}(1+s\frac{x^2}{4l^2})^4 \frac{\partial r_b}{\partial x^\mu}\frac{\partial r_c}{\partial x^\nu}\frac{\partial r_d}{\partial x^\lambda}\frac{\partial r_e}{\partial x^\rho}(r_a \e^{abcde})
\label{laltu60}
\eer

      Now the explicit expressions for the anomaly are  identified with the minimum of effort.
Indeed, using (\ref{bibhas10}) and (\ref{laltu60}), the ABJ anomaly is projected as,
\be
\frac{1}{16\pi^2} \e_{\mu\nu\lambda\rho} F^{\mu\nu}F^{\lambda\rho} = s\frac{1}{16\pi^2 l^5} (1+s\frac{x^2}{4l^2})^{-4}r^a r_f r_i \e_{abcde} \hat{F}^{fbc}\hat{F}^{ide}
\label{laltu61}
\ee
The weight factors cancel out from both sides of the projected
anomaly equation (\ref{anomaly4}) and we obtain, using (\ref{m9}) and (\ref{laltu61}),
\be
 r_a L^{ab} \hat j_{b5} = \frac{ig^2}{16\pi^2 l^3} r^a r_f r_i \e_{abcde} \hat{F}^{fbc}\hat{F}^{ide}
\label{180}
\ee
It is also possible to rewrite the anomaly expression in a form that resembles the expression on the hypersphere \cite{banerjee,donnell2}. To do this we have to exploit the identity,
\be
\e_{abcde} \hat{F}^{abc} = s\frac{3}{l^2}r^a r_f \e_{abcde} \hat{F}^{fbc}
\label{laltu62}
\ee 
Then the A(dS) anomaly equation reduces to,
\be
 r_a L^{ab} \hat j_{b5} = s {ig^2\over {48\pi^2 l}}r_a\e_{bcdef}\hat F^{abc}\hat F^{def}
\label{m12}
\ee

This is the desired anomalous current divergence equation in the A(dS) space which has a close resemblance with the corresponding equation on the hypersphere. It is 
the exact analogue of the ABJ-anomaly equation on the flat space.

      There is another way in which the anomaly equation can be expressed. To see this, observe that the projection (\ref{laltu21}) for the current corresponds to $n=2$ in the general formula (\ref{ma11}). Hence  the identity (\ref{ma12}) holds and we obtain,
\be
r_c r_a L^{ab} \hat j_{b5} = sl^2(L_{cb}\hat{j}^{b5}-\hat{j}_{c5})
\label{m12a}
\ee
 Thus the anomaly equation (\ref{m12}) takes the form given in (\ref{m12d}), thereby completing our proof of equivalence.
Compatibility between the two forms (\ref{m12}) and (\ref{m12d}) is easily established by
contracting the latter with $r^g$ and using the transversality  of the current ($r^a \hat{j}_{a5} =0$).

  The normal Ward identity for the vector current is obtained by setting the right hand side of 
either (\ref{m12}) or (\ref{m12d}) equal to zero.

                It is known that  on the flat space, it is feasible to redefine the current so that the anomaly vanishes. In that case, however, the current is no longer gauge invariant. This compensating term is given by,
\be
X^\mu = \frac{ig^2}{8\pi^2} \e{^\mu}{^\nu}{^\lambda}{^\rho} A_\nu F{_\lambda}{_\rho}
\label{bibhas18}
\ee
such that  $\partial_\mu J^{\mu 5} = 0$, where,
\be
J^{\mu 5} = j^{\mu 5} -X^\mu.
\ee 
Observe that $J^{\mu 5}$ is not gauge invariant due to the presence of $X^\mu$.

           The same phenomenon also occurs on the A(dS) space. Here the compensating piece is obtained from a projection of (\ref{bibhas18}),
\be
\hat{X}_a = (1+s\frac{x^2}{4l^2})^2 K{_a}{^\mu} X_\mu
\ee
Since,
\ber
r_a L^{ab}\hat{X}_b &=& s l^2 (1+s\frac{x^2}{4l^2})^4 \partial_\mu X^\mu
\nonumber
\\&=& s\frac{ig^2}{48 \pi^2 l} r_a \e_{bcdef} \hat{F}^{abc} \hat{F}^{def}
\eer
we observe that the modified current,
\ber
\hat{J}_{a5}&=& \hat{j}_{a5} - \hat{X}_a
\nonumber
\\&=&(1+s\frac{x^2}{4l^2})^2 K{_a}{^\mu} J_{\mu5}
\label{laltu32}
\eer
is anomaly free,
i.e. $r_a L^{ab} \hat{J}_{b5} = 0$. Also note that the anomaly free currents are mapped in the same way as the anomalous ones. The transversality condition  $r^a \hat{J}_{a5} = 0$ is obviously satisfied by (\ref{laltu32}). Expectedly, the current $\hat{J}_{a5}$ is not gauge invariant.

       It is straightforward to extend this calculation for arbitrary $D=2n$ dimensions \footnote{For simplicity, the coupling factor $g$ is not included.}. The flat space expression (\ref{anomaly4}) is known to be generalised as \cite{BZ,HRP},
\be
\partial_\mu j^{\mu 5}=\frac{2i}{(4\pi)^n n!}\e_{\mu_1 \mu_2........\mu_{2n}}F^{\mu_1\mu_2}F^{\mu_3\mu_4}......F^{\mu_{2n-1}\mu_{2n}}
\label{1}
\ee
The map for the Levi-Civita tensor is the generalised version of (\ref{laltu60}),
\be
\e_{\mu_1\mu_2.....\mu_{2n}}=\frac{s}{l}(1+s\frac{x^2}{4l^2})^{-2n}r_a K_{a_1\mu_1}K_{a_2\mu_2}.....K_{a_{2n}\mu_{2n}}\e^{aa_1a_2.....a_{2n}}
\label{2}
\ee
while the map for the field tensor is given by (\ref{bibhas10}). Using these mappings the projected expression for the anomaly is,
\ber
&&\frac{2}{(4\pi)^nn!}\e_{\mu_1\mu_2......\mu_{2n}}F^{\mu_1\mu_2}.......F^{\mu_{2n-1}\mu_{2n}}=\frac{2}{(4\pi)^nn!}\frac{s}{l^{2n+1}} (1+s\frac{x^2}{4l^2})^{-2n}
\nonumber
\\
&&r^a r_{a_1}r_{a_2}......r_{a_n}\e_{ab_1b_2.....b_{2n}}
\hat{F}^{a_1b_1b_2}\hat{F}^{a_2b_3b_4}.......\hat{F}^{a_nb_{2n-1}b_{2n}}
\label{3}
\eer
Next, the projection of the L.H.S. of (\ref{anomaly4}) has to be found. Using (\ref{new}) and (\ref{133}) we obtain,
\be
r_a L^{ab}\hat{j}{_b}{^5}=sl^2(1+s\frac{x^2}{4l^2})^{2n}\partial^\mu j{_\mu}{^5}
\label{4}
\ee
Finally, exploiting (\ref{3}) and (\ref{4}) the projected form of (\ref{1}) is derived,
\be
r_aL^{ab}\hat{j}{_b}{^5}=\frac{2i}{(4\pi)^nn!}\frac{1}{l^{2n-1}}r^ar_{a_1}r_{a_2}..........r_{a_n}\e_{ab_1b_2.....b_{2n}}\hat{F}^{a_1b_1b_2}\hat{F}^{a_2b_3b_4}......\hat{F}^{a_nb_{2n-1}b_{2n}}
\label{5}
\ee
Following the same steps employed for obtaining (\ref{m12d}) from (\ref{180}), the above anomaly equation is expressed as,
\be
\hat{j}_{g5}-L_{gb}\hat{j}^{b5}=-\frac{2i}{3(4\pi)^nn!l^{2n-1}}r_gr_a\e_{a_1a_2.....a_{2n+1}}\hat{F}^{aa_1a_2}\hat{F}^{a_3a_4a_5}.......\hat{F}^{a_{2n-1}a_{2n}a_{2n+1}}
\label{6}
\ee

\subsection{Non-abelian Chiral Anomalies}

         Now we will discuss about the non-abelian chiral anomaly of spin one-half fermions.  As is well known \cite{BZ,HRP,fuji,bertl}, there are two types of anomaly  on the flat Minkowski space : the covariant anomaly and the consistent anomaly. The covariant anomaly, as its name implies, transforms covariantly under the gauge transformation. The consistent anomaly, on the other hand, is the one that satisfies the Wess-Zumino consistency condition. Just as the covariant anomaly does not satisfy this condition, the consistent anomaly does not transform covariantly.

    The covariant anomaly is given by,
\ber
(D_\mu j^{\mu})^{(\alpha)}_{(covariant)} &=& (\partial_\mu j^\mu)^{(\alpha)} - i[A_\mu,j^\mu]^{(\alpha)}
\nonumber
\\
\nonumber
\\&=&\frac{i}{32 \pi^2} \e^{\mu\nu\rho\sigma} Tr.\{\lambda^\alpha F_{\mu\nu} F_{\rho\sigma}\}
\label{field2}
\eer
where $\lambda^\alpha$ are the symmetry matrices and $(j^\mu)^{(\alpha)}$ is the chiral current,
\be
(j_\mu)^{(\alpha)} = \bar{\Psi} \lambda^\alpha \gamma_\mu \frac{1+\gamma_5}{2}\Psi
\label{field1}
\ee
The result (\ref{field2}) has been obtained by various methods \cite{BZ,HRP,fuji,bertl}, all of which basically rely on regularising the current (\ref{field1}) in a covariant manner.

          We shall now stereographically project (\ref{field2}) to obtain the covariant anomaly on the A(dS) space. The map for the current (\ref{field1}) is similar to (\ref{new9}) and is given by,
\be
(\hat{j}_a)^{(\alpha)}=(1+s\frac{x^2}{4l^2})^2 K{_a}{^\mu}(j_\mu)^{(\alpha)}
\ee
Now, using (\ref{new1}) and following steps similar to the abelian case, we find,
\be
(r^a \hat{\cal{L}}_{ab}\hat{j}^{b})^{(\alpha)} = sl^2 (1+s\frac{x^2}{4l^2})^4 (D_\mu j^{\mu})^{(\alpha)}
\label{field3}
\ee
which is the non-abelian version of (\ref{m9}).

    The R.H.S. of (\ref{field2}) is obtained by a straightforward generalisation of (\ref{laltu61}),
\be
\frac{1}{32 \pi^2} \e^{\mu\nu\rho\sigma} Tr.\{\lambda^\alpha F_{\mu\nu} F_{\rho\sigma}\}= \frac{s}{32\pi^2 l^5}(1+s\frac{x^2}{4l^2})^{-4}r^ar_fr_i\e_{abcde}Tr.\{\lambda^\alpha \hat{F}^{fbc}\hat{F}^{ide}\}
\label{field4}      
\ee
Hence, the covariant anomaly on A(dS) space is given by,
\be
(r^f\hat{\cal{L}}_{fg}\hat{j}^{g})^{(\alpha)}_{(covariant)} = \frac{is}{96\pi^2 l} r_a \e_{bcdef} Tr.\{\lambda^\alpha \hat{F}^{abc}\hat{F}^{def}\}
\ee
which follows on exploiting (\ref{field2}), (\ref{field3}), (\ref{field4}) and the identity (\ref{laltu62}).

    Next, the consistent anomaly is considered. On the flat Minkowski space this is given by,
\be
(D_\mu j^{\mu})^{(\alpha)}_{(consistent)} = \frac{i}{24 \pi^2} \e^{\mu\nu\rho\sigma} Tr.\{\lambda^\alpha \partial_\mu (A_\nu \partial_\rho A_\sigma - \frac{i}{2} A_\nu A_\rho A_\sigma)\}
\ee
In order to project this equation it is convenient to recast it in the following form,
\ber
&(D_\mu j^{\mu })_{(consistent)}={\cal{A}}_{(consistent)}^{(\alpha)}& 
\nonumber
\\
\nonumber
\\= &\frac{i}{96 \pi^2}\e^{\mu\nu\rho\sigma} Tr.\{\lambda^\alpha (F_{\mu\nu} F_{\rho\sigma} + i F_{\mu\nu}A_\rho A_\sigma + i A_\mu A_\nu F_{\rho\sigma} - i A_\nu F_{\mu\rho}A_\sigma)\}&
\label{field5}
\eer
where the definition (\ref{v2}) of field tensor $F_{\mu\nu}$ has been used.
Using the inverse maps for $\e^{\mu\nu\rho\sigma}$ (\ref{laltu60}), field tensor (\ref{bibhas10}) and the vector field (\ref{bibhas9}), the stereographic projection of the consistent anomaly (\ref{field5}) on A(dS) space is,
\ber
({\cal{A}})_{(consistent)}^{(\alpha)} = \frac{i}{96\pi^2 l^3} (1+s\frac{x^2}{4l^2})^{-4} \e_{abcde} r^a r_f [\frac{s}{l^2}r_i Tr.\{\lambda^\alpha \hat{F}^{fbc}\hat{F}^{ide}\} 
\nonumber
\\
\nonumber
\\+ i Tr.\{\lambda^\alpha (\hat{F}^{fbc}\hat{A}^d \hat{A}^e + \hat{A}^c \hat{A}^d \hat{F}^{fbc} - \hat{A}^c \hat{F}^{fbd}\hat{A}^e)\}]
\eer
Now using the identity (\ref{laltu62}) one can show,
\ber
({\cal{A}})_{(consistent)}^{(\alpha)}  = \frac{i}{288\pi^2 l} (1+s\frac{x^2}{4l^2})^{-4}  [\frac{1}{l^2}r_a\e_{bcdef} Tr.\{\lambda^\alpha \hat{F}^{abc}\hat{F}^{def}\} 
\nonumber
\\
\nonumber
\\+ is\e_{abcde} Tr.\{\lambda^\alpha (\hat{F}^{abc}\hat{A}^d \hat{A}^e + \hat{A}^c \hat{A}^d \hat{F}^{abc} - \hat{A}^c \hat{F}^{abd}\hat{A}^e)\}]
\eer
Therefore, the final equation for the consistent anomaly  on the A(dS) space is,
\ber
(r^f \hat{\cal{L}}_{fg}\hat{j}^{b})_{(consistent)}^{(\alpha)} = s\frac{il}{288\pi^2}[\frac{1}{l^2}r_a\e_{bcdef}Tr.\{\lambda^\alpha \hat{F}^{abc}\hat{F}^{def}\}
\nonumber
\\
\nonumber 
\\+ is \e_{abcde} Tr.\{\lambda^\alpha(\hat{F}^{abc}\hat{A}^d \hat{A}^e + \hat{A}^d \hat{A}^e \hat{F}^{abc} + \hat{A}^d \hat{F}^{abc}\hat{A}^e)\}]
\eer

    It is known that, on the flat Minkowski space, the covariant  and consistent currents are related by a local counterterm given by,
\ber
(X^\mu)^{(\alpha)} = (j^\mu)_{(consistent)}^{(\alpha)}-(j^\mu)_{(covariant)}^{(\alpha)}
\nonumber
\\
\nonumber 
\\=-\frac{i}{48 \pi^2}\e^{\mu\nu\rho\sigma} Tr.\{\lambda^\alpha(A_\nu F_{\rho\sigma}+F_{\rho\sigma}A_\nu + i A_\nu A_\rho A_\sigma)\}
\label{field7}
\eer
As was shown in \cite{HRP}, this ambiguity in the current is such that it does not affect the effective action. This is a consequence of the fact that,
\be
(A^\mu)^{(\alpha)} (X_\mu)^{(\alpha)}= 0
\label{field6} 
\ee
From (\ref{field7}) we observe that the covariant and consistent anomalies on flat space are related by,
\be
(D_\mu j^{\mu})^{(\alpha)}_{(covariant)} = (D_\mu j^{\mu})^{(\alpha)}_{(consistent)} - (D_\mu X^\mu)^{(\alpha)}
\label{baba}
\ee

     The above analysis is now carried out on the A(dS) space. The projection of (\ref{field7}) yields,
\ber
&(\hat{X}_b)^{(\alpha)} = (1+s\frac{x^2}{4l^2})^2 K_{b\mu} (X^\mu)^{(\alpha)}&
\nonumber
\\
\nonumber
\\&=-\frac{i}{48\pi^2 l}r^a\e_{abcde}[\frac{1}{l^2}r_iTr.\{\lambda^\alpha(\hat{A}^c\hat{F}^{ide}+\hat{F}^{ide}\hat{A}^c)\} +isTr.\{\lambda^\alpha\hat{A}^c\hat{A}^d\hat{A}^e\}]&
\label{field8}
\eer
Note that the condition analogous to (\ref{field6}) is,
\be
(\hat{A}^a)^{(\alpha)} (\hat{X}_a)^{(\alpha)} = 0
\ee
which is obviously satisfied by (\ref{field8}).
Now, using the definition (\ref{new1}) of `covariantised angular momentum' and (\ref{field8}), we have, 
\ber
(r^f\hat{\cal{L}}_{fg}\hat{X}^g)^{(\alpha)} = -s\frac{il}{288\pi^2}[\frac{2}{l^2}r_a \e_{bcdef} Tr.\{\lambda^\alpha \hat{F}^{abc}\hat{F}^{def}\} 
\nonumber
\\
\nonumber
\\-is \e_{abcde}Tr.\{\lambda^\alpha(\hat{F}^{abc}\hat{A}^d\hat{A}^e + \hat{A}^d\hat{A}^e\hat{F}^{abc} + \hat{A}^d\hat{F}^{abc}\hat{A}^e)\}]
\eer
 
Exploiting the expressions for the consistent anomaly, covariant anomaly and the above relation  we have,
\be
(r^f\hat{\cal{L}}_{fg} \hat{j}^{g})^{(\alpha)}_{(covariant)} = (r^f\hat{\cal{L}}_{fg} \hat{j}^{g})^{(\alpha)}_{(consistent)} - (r^f\hat{\cal{L}}_{fg} \hat{X}^g)^{(\alpha)}
\ee
which is the A(dS) space analogue of (\ref{baba}). Here, $(r^f\hat{\cal{L}}_{fg} \hat{X}^g)^{(\alpha)}$ plays the role of the local counterterm for the anomaly on the A(dS) space.

\section{Duality Symmetry}

The well known electric-magnetic duality symmetry swapping 
field equations with the Bianchi identity in flat space has an exact counterpart on
the A(dS) hyperboloid. To see this it is essential to introduce the dual field 
tensor that enters the Bianchi identity. The dual tensor is defined by,
\be
\tilde F_{ab} = -\frac{1}{6}\epsilon_{abcde}\hat F^{cde}
\label{d1}
\ee

Using (\ref{v11}) and (\ref{m5}) together with the properties of the Killing vectors the 
dual on the A(dS) space is expressed in terms of the dual on the flat space as,
\be
\tilde F_{ab} = -sl K_a^\lambda K_b^\rho \tilde F_{\lambda\rho}
\label{d2}
\ee
where the flat space dual is given by,
\be
\tilde F_{\lambda\rho} = \frac{1}{2}\epsilon_{\lambda\rho\mu\nu} F^{\mu\nu}
\label{d3}
\ee
Inversion of (\ref{d2}) yields,
\ber
\tilde F_{\mu\sigma}&=& -s \frac{1}{l}(1+s\frac{x^2}{4l^2})^{-4} K_{a\mu} K_{b\sigma} \tilde F^{}ab 
\nonumber
\\
\nonumber
\\&=& -s\frac{1}{l} \frac{\partial{r_a}}{\partial{x^\mu}}\frac{\partial{r_b}}{\partial{x^\sigma}} \tilde F^{ab}
\eer
where we have used (\ref{deser}) to obtain the final result. Apart from a dimensional scale the mapping is exactly identical to that of a second rank tensor given in (\ref{bibhas21}).

The Bianchi identity on the A(dS) space is then given by,
\be
r_a L^{ab} \tilde F_{bc} = 0
\label{d4}
\ee
This is confirmed by a direct calculation. Alternatively, it becomes transparent by projecting it
on the flat space by means of Killing vectors. Using the basic definitions and the
identity,
\be
K_b^\rho \p_\rho(K^{b\mu} K_c^\nu)\tilde F_{\mu\nu} = 0
\label{zc2}
\ee
we obtain,
\be
r_a L^{ab} \tilde F_{bc} = -l^3 K^{b\mu}K_b^\lambda K_c^\rho \p_\mu\tilde F_{\lambda\rho}
\label{d5}
\ee
Finally, exploiting  (\ref{k1}) we get the desired projection,
\be
r_a L^{ab} \tilde F_{bc} = -l^3 \Big(1+s\frac{x^2}{4l^2}\Big)^2 K_c^\rho \p^\lambda\tilde F_{\lambda\rho}
\label{d6}
\ee
which vanishes since $\p^\lambda\tilde F_{\lambda\rho}=0$.

Now the abelian equation of motion following from a variation of the action (\ref{v20})
is given by,
\be
L_{ab} \hat F^{abc} = 0
\label{d7}
\ee

The duality transformation is next discussed. Analogous to the flat space rule, 
$\tilde F\rightarrow F; F\rightarrow -\tilde F$ the duality map here is provided by,
\be
\tilde F_{ab} \rightarrow \frac{r^c}{l} \hat F_{abc}\ \ \ ;\ \ \ \frac{r^c}{l}\hat F_{abc}\rightarrow -\tilde F_{ab}
\label{d8}
\ee
It is easy to check the consistency of this map. The inverse of (\ref{d1}) yields,
\be
\hat F_{abc} = s\frac{1}{2}\epsilon_{abcde}\tilde F^{de}
\label{d9}
\ee
while,
\be
\hat{F}^{abc} = -\frac{1}{2} \epsilon^{abcde} \tilde F_{de}
\label{bibhas50}
\ee
Under the first of the maps in (\ref{d8}), the above relation is transformed as,
\be
\hat F_{abc}\rightarrow  -s\frac{1}{4l}r_f \epsilon_{abcde} \epsilon^{defgh} \tilde{F}_{gh} =  -s\frac{1}{l}\Big(r_a \tilde F_{bc} +r_b \tilde F_{ca}+ r_c \tilde F_{ab}\Big)
\label{d10}
\ee
where use was made of (\ref{bibhas50}) at an intermediate step. Contracting the above map by $r^c$
immediately leads to the second relation in (\ref{d8}).

Likewise, under the map (\ref{d10}), the dual field (\ref{d1}) transforms as,
\be
\tilde{F}_{ab} \rightarrow s\frac{1}{2l} \epsilon_{abcde} r^c \tilde{F}^{de}
\ee
Substituting the expression (\ref{d1}) for $\tilde{F}^{de}$ we reproduce the first of the maps given in (\ref{d8}).

     Now the effect of the duality map on the equation of motion (\ref{d7}) is considered. Using 
(\ref{d10}) and the correspondence (\ref{d2}) along with the identity (\ref{zc2}) we find,
\be
\frac{1}{2}L_{ab} \hat F^{abc} \rightarrow sl^2\Big(1+s\frac{x^2}{4l^2}\Big)^2 K^{c\rho}\p^\mu \tilde F_{\mu\rho}
\label{d11}
\ee
Finally, using (\ref{d6}) we obtain the cherished mapping,
\be
\frac{1}{2}L_{ab} \hat F^{abc} \rightarrow -s\frac{1}{l} r_a L^{ab} \tilde F_{bc}
\label{d11}
\ee
showing how the equation of motion passes over to the Bianchi identity. Likewise the 
other map swaps the Bianchi identity to the equation of motion,
\be
s\frac{1}{l} r_a L^{ab} \tilde{F}_{bc} \rightarrow \frac{1}{2} L_{ab} \hat{F}^{abc}
\ee

It is feasible to perform a continuous $SO(2)$ duality rotations through an angle $\theta$. The relevant transformations are
then given by,
\ber
\frac{r^c}{l} \hat{F}_{abc}' &=& \mbox{Cos} \theta \ \ \frac{r^c}{l} \hat{F}_{abc} - \mbox{Sin} \theta\ \ \tilde F_{ab} \\
\nonumber
\\
\tilde F_{ab}' &=& \mbox{Sin} \theta \ \ \frac{r^c}{l} \hat{F}_{abc} + \mbox{Cos} \theta \ \ \tilde F_{ab}
\label{d12}
\eer
This mixes the equation of motion and the Bianchi identity in the following way,
\ber
\frac{1}{2} L^{ab}\hat{F}_{abc}' &=& \mbox{Cos} \theta \ \ \frac{1}{2} L^{ab} \hat{F}_{abc} - \mbox{Sin} \theta\ \ s\frac{1}{l}r_aL^{ab}\tilde F_{bc} \\
\nonumber
\\
s\frac{1}{l}r_a L^{ab} \tilde F_{bc}' &=& \mbox{Sin} \theta \ \ \frac{1}{2}L^{ab} \hat{F}_{abc} + \mbox{Cos} \theta \ \ s\frac{1}{l}r_a L^{ab}\tilde F_{bc}
\label{d12}
\eer

The discrete duality transformation corresponds to $\theta=\frac{\pi}{2}$.

\section{ Formulation of Antisymmetric Tensor Gauge Theory}

\bigskip

          Our analysis can be extended to include higher rank tensor gauge theories. Some typical examples are the linearised version of gravity which uses a symmetric second rank tensor or the p-form gauge theories employing antisymmetric tensor fields.

         In this section we discuss our formulation for the 
second rank antisymmetric tensor gauge theory. Also, there are some features
which distinguish it from the analysis for the  vector gauge theory.
The extension for higher forms 
is obvious. Both abelian and nonabelian theories will be considered. To set
up the formulation it is convenient to begin with the abelian case which can
be subsequently generalised to the nonabelian version. The action for a free
2-form gauge theory in flat four-dimensional  Minkowski space is given by \cite{KR},
\be
S= -{1\over {12}}\int d^4 x F^{\mu\nu\rho}F_{\mu\nu\rho}
\label{t1}
\ee
where the field strength is defined in terms of the basic field as,
\be
 F_{\mu\nu\rho}= \p_\mu B_{\nu\rho}+ \p_\nu B_{\rho\mu}+ \p_\rho B_{\mu\nu}
\label{t2}
\ee
The infinitesimal gauge symmetry is given by the transformation,
\be
\d B_{\mu\nu} = \p_\mu \Lambda_\nu -\p_\nu \Lambda_\mu
\label{t3}
\ee
which is reducible since it trivialises for the choice $\Lambda_\mu = \p_\mu\lambda$.

It is sometimes useful to express the action (or the lagrangian) in a first order form
by introducing an extra field,
\be
{\cal L}= -{1\over {8}}\e_{\mu\nu\rho\sigma} F^{\mu\nu}B^{\rho\sigma}+ {1\over 8}A^\mu A_\mu
\label{t4}
\ee
where the $B\wedge F$ term involves the field tensor,
\be
F_{\mu\nu}=\p_\mu A_\nu - \p_\nu A_\mu
\label{t5}
\ee
Eliminating the auxiliary $A_\mu$ field by using its equation of motion, the previous
form (\ref{t1}) is reproduced. The gauge symmetry is given by (\ref{t3}) together with
$\d A_\mu=0$. The first order form is ideal for analysing the nonabelian theory.

To express the theory on the A(dS) pseudosphere, the mapping of the tensor field is first given.
From the previous analysis, it is simply given by,
\be
\hat B_{ab} = K_a^\mu K_b^\nu B_{\mu\nu} 
\label{t9}
\ee
and satisfies the transversality condition,
\be
r^a \hat{B}_{ab} = r^b \hat{B}_{ab} = 0.
\label{bibhas20}
\ee
The tensor field with the latin indices is defined on the pseudosphere while those with
the greek symbols are the usual one on the flat space.
This is written in component notation by using the explicit form  for the Killing vectors 
given in (\ref{s18a}) and (\ref{s18b}),
\be
\hat B_{\m\n} =\Big(1 +s\frac{x^2}{4l^2}\Big)\Big((1+s\frac{x^2}{4l^2}) B_{\m\n} -s \frac{x^\rho x_\n}{2l^2} B_{\m\rho}
-s\frac{x^\rho x_\m}{2l^2} B_{\rho\n}\Big)
\label{t10}
\ee
and,
\be
\hat B_{\mu 4} = \frac{1}{l}\Big(1+s\frac{x^2}{4l^2}\Big) x^\rho B_{\m\rho}
\label{t11}
\ee
These are the analogues of (\ref{newmap}). The inverse relation is given by,
\be
\Big(1+s\frac{x^2}{4l^2}\Big)^4 B^{\m\n} = K_a^\m K_b^\n \hat B^{ab}
\label{t12}
\ee
which may also be put in the form,
\be
\Big(1+s\frac{x^2}{4l^2}\Big)^2 B_{\m\n} =  \hat B_{\m\n} -s\frac{x_\m \hat B_{\n 4}}{2l}+s \frac{x_\n \hat B_{\m 4}}{2l}
\label{t13}
\ee
which is the direct analogue of (\ref{inverse}). Moreover using (\ref{deser}), the map (\ref{t12}) may also be expressed in a more transparant form as,
\be
B_{\mu\nu} = \frac{\partial{r_a}}{\partial{x^\mu}}\frac{\partial{r_b}}{\partial{x^\nu}} \hat{B}^{ab}
\ee
which is just the result (\ref{bibhas21}) for a second rank tensor.

            Next, the gauge transformations are discussed. From (\ref{t3}), the defining relation (\ref{t9}) and the 
angular momentum operator (\ref{v8}), infinitesimal transformations are given by,
\be
\d \hat B_{ab}= s\frac{r^c}{l^2}\Big(K_b^\m L_{ca} -K_a^\m L_{cb}\Big)\Lambda_\m
\label{t14}
\ee
This is consistant with $r^a \delta\hat{B}_{ab} = 0$ which is imposed by (\ref{bibhas20}).
In this form the expression is not manifestly covariant. This may be contrasted with
(\ref{v10}) which has this desirable feature. The point is that an appropriate map
of the gauge parameter is necessary. In the previous example the gauge parameter was a scalar
which retained its form. Here, since it is a vector, the required map is provided by a relation
like (\ref{k}), so that,
\be
\hat\Lambda_a = K_a^\mu \Lambda_\mu
\label{t14.1}
\ee
Pushing the Killing vectors through the angular momentum operator and using the above 
map yields, after some simplifications,
\be
\d \hat B_{ab}= s\frac{1}{l^2}\Big[ r^c\Big( L_{ca}\hat\Lambda_b - L_{cb}\hat\Lambda_a \Big)
- r_a\hat\Lambda_b + r_b\hat\Lambda_a\Big]
\label{t14.3}
\ee

It is also reassuring to note that (\ref{t14.3}) manifests the reducibility of the
gauge transformations. Since $\Lambda_\mu=\p_\mu\lambda$ leads to a trivial gauge
transformation in 
flat space, it follows from (\ref{t14.1}) that the corresponding feature should be present
in the pseudospherical formulation when,
\be
\hat\Lambda_a = s\frac{1}{l^2} r^c L_{ca}\lambda
\label{t14.5}
\ee
It is easy to check that with this choice, the gauge transformation (\ref{t14.3}) 
trivialises; i.e. $\d \hat B_{ab}=0$.

The field tensor on the pseudosphere is constructed from the usual one given in (\ref{t2}).
Since the Killing vectors play the role of the metric in connecting the two surfaces, this
expression is given by a natural extension of (\ref{v11}),
\be
\hat F_{abcd}= \Big(r_a K_b^\mu K_c^\nu K_d^\rho+r_b K_c^\mu K_a^\nu K_d^\rho + r_c K_d^\mu K_a^\nu
K_b^\rho +r_d K_a^\mu K_c^\nu K_b^\rho\Big)F_{\mu\nu\rho}
\label{t15}
\ee
Note that cyclic permutations have to taken carefully since there is an even number of
indices. The inverse mapping is provided by,
\ber
F_{\mu\nu\rho} &=& s\frac{1}{l^2} (1+s\frac{x^2}{4l^2})^{-6} K_{b\mu} K_{c\nu} K_{d\rho} (r_a \hat{F}^{abcd}) 
\nonumber
\\
\nonumber
\\&=& s\frac{1}{l^2} \frac{\partial{r_b}}{\partial{x^\mu}}\frac{\partial{r_c}}{\partial{x^\nu}}\frac{\partial{r_d}}{\partial{x^\rho}}(r_a \hat{F}^{abcd})
\label{bibhas23}
\eer
This is a particular case of the general result (\ref{bibhas22}).

         In terms of the basic variables, the field tensor is known to be expressed as,
\be
\hat F_{abcd}= \Big(L_{ab}\hat B_{cd} +L_{bc}\hat B_{ad} + L_{bd}\hat B_{ca} + L_{ca}\hat B_{bd} +
L_{da}\hat B_{cb}+
L_{cd}\hat B_{ab}\Big)
\label{t16}
\ee

To show that (\ref{t15}) is 
equivalent to (\ref{t16}), the same strategy as before, is 
adopted. Using the definition of the angular 
momentum (\ref{v8}), (\ref{t16}) is simplified
as,
\be
\hat F_{abcd}=\Big(r_a K_b^\mu - r_b K_a^\mu\Big)\p_\mu\Big( K_c^\nu K_d^\sigma B_{\nu\sigma} \Big)
+............
\label{t17}
\ee
where the carets denote the inclusion of other similar 
(cyclically permuted) terms. Now there are two types
of contributions. Those where the derivatives act on the Killing vectors and those where
they act on the fields. The first class of terms cancel out as a consequence of an identity
that is an extension of (\ref{v14}). The other class combines to reproduce (\ref{t15}).

The action on the A(dS) pseudosphere is now obtained by first taking a repeated product of the field
tensor (\ref{t15}). Using the properties of the Killing vectors, this yields,
\be
\hat F_{abcd}\hat F^{abcd}= 4sl^2\Big(1+s\frac{x^2}{4l^2}\Big)^6 F_{\mu\nu\rho}F^{\mu\nu\rho}
\label{t18}
\ee
From the definition of the flat space action (\ref{t1}) and the volume element (\ref{conformal}),
it follows that the above identification leads to the pseudospherical action,
\be
S_\Omega =-s\frac{1}{48 l^2}\int d\Omega \Big(1+s\frac{x^2}{4l^2}\Big)^{-2}\hat F_{abcd}\hat F^{abcd}
\label{t19}
\ee

Thus, up to a conformal factor, the corresponding lagrangian is given by,
\be
{\cal L}_\Omega =-s\frac{1}{48 l^2}\hat F_{abcd}\hat F^{abcd}
\label{t20}
\ee

By its very construction this lagrangian would be invariant under the gauge transformation
(\ref{t14.3}). There is however another type of gauge symmetry 
which does not seem to have any analogue in the flat space. 
To envisage such a  possibility, consider a  transformation of the
type {\footnote{Recently such a transformation was considered on the 
hypersphere \cite{M, banerjee}}},
\be
\delta \hat B_{ab}=L_{ab}\lambda
\label{t21}
\ee
which  could be a meaningful gauge symmetry
operation on the A(dS) space. However, in flat space, it
leads to a trivial gauge transformation. To see this explicitly, consider the effect of
(\ref{t21}) on (\ref{t12}),
\be
\Big(1+s\frac{x^2}{4l^2}\Big)^4 \delta B^{\m\n} = K_a^\m K_b^\n L_{ab}\lambda
\label{t22}
\ee
Inserting the expression for the angular momentum from (\ref{v8}) and
exploiting the transversality (\ref{s8}) of the Killing vectors, it follows that,
\be
 \delta B_{\m\n} = 0
\label{t23}
\ee
thereby proving  the statement. To reveal that (\ref{t21})
indeed leaves the lagrangian (\ref{t20}) invariant, it is desirable 
to recast it in the form,
\be
{\cal L}_\Omega =-s\frac{1}{32l^2}\hat \Sigma_{a}\hat \Sigma^{a}
\label{t24}
\ee
where,
\be
\hat \Sigma_a = \e_{abcde}L^{bc}\hat B^{de}
\label{t25}
\ee
Under the gauge transformation (\ref{t21}), a simple algebra shows that $\delta \hat \Sigma_a = 0 $ and hence 
the lagrangian remains invariant.

The inclusion of a nonabelian gauge group is feasible. Results follow logically
from the abelian theory with suitable insertion of the nonabelian indices. As
remarked earlier it is useful to consider the first order form (\ref{t4}). The
lagrangian is given by its straightforward generalisation ,
where the nonabelian field strength has already been defined in (\ref{v2}). It is
gauge invariant under the nonabelian generalisation of (\ref{t3}) with the 
ordinary derivatives replaced by the covariant derivatives with respect to the potential 
$A_\mu$, and $\d A_\mu = 0$. By the help of our equations it is possible to
project this lagrangian on the A(dS) space. For instance, the corresponding gauge 
transformations look like,
\ber
\d \hat B_{ab} &=& s\frac{1}{l^2} \Big[r^c\Big( L_{ca}\hat\Lambda_b - L_{cb}\hat\Lambda_a \Big)
- r_a\hat\Lambda_b + r_b\hat\Lambda_a\Big] -i [\hat A_a ,  \hat\Lambda_b]+i[\hat{A}_b,\hat{\Lambda}_a]
\nonumber
\\
\nonumber
\\&=&\frac{s}{l^2}[r^c(\hat{\cal{L}}_{ca}\hat{\Lambda}_b - \hat{\cal{L}}_{cb}\hat\Lambda_a)-r_a\hat\Lambda_b+r_b\hat\Lambda_a]
\label{t27}
\eer
and so on. Expectedly, the ordinary angular momentum operator gets replaced by its covariantised version.

Matter fields may be likewise defined. The fermion current $j_{\mu\nu}$ will
be defined just as the two form field,
\be
\hat{j}_{ab} = K_{a\mu} K_{b\nu} j^{\mu\nu}
\ee
while the inverse relation is given by,
\ber
j_{\mu\nu} &=& (1+s\frac{x^2}{4l^2})^{-4} K_{a\mu} K_{b\nu} \hat{j}^{ab}
\nonumber
\\
\nonumber
\\&=& \frac{\partial{r_a}}{\partial{x^\mu}}\frac{\partial{r_b}}{\partial{x^\nu}} \hat{j}^{ab}
\eer
It is simple to verify that this map preserves the form invariance of the interaction,
\be
\int d^4x (j_{\mu\nu} B^{\mu\nu}) = \int d\Omega (\hat j_{ab} \hat B^{ab})
\label{t28}
\ee
quite akin to (\ref{m1}).

\section {Zero Curvature Limit}
\bigskip

The null curvature limit (which is also equivalent to a vanishing cosmological constant) is obtained
by setting $l\rightarrow \infty$. Then the A(dS) group contracts to the 
Poincare group so that the field theory on the A(dS) space should contract to the corresponding theory
on the flat Minkwski space. This  is shown  very conveniently in the present formalism 
using Killing vectors. The example of Yang Mills theory with sources will be considered.

The equation of motion in the A(dS) space  obtained by varying the action composed of the pieces
(\ref{v20}) and (\ref{m1}) is found to be,
\be
s\frac{1}{2 l^2}\hat{\cal L}_{ab}\hat F^{abc} + \hat j^c = 0
\label{zc1}
\ee

The operator appearing in the above equation is now mapped to the flat space. The mapping
for the usual angular
momentum part is first derived,
\be
L_{ab} \hat F^{abc} = 2 r_a K_b^\mu \p_\mu\Big([r^a K^{b\nu} K^{c \rho} +c.p.] F_{\nu\rho}\Big)
\label{zero}
\ee
Using the transversality condition and the identities among the Killing vectors it is seen that the
only nonvanishing contribution comes from the action of the derivative on the field tensor yielding,
\be
 L_{ab} \hat F^{abc} = 2s l^2\Big(1+s\frac{x^2}{4l^2}\Big)^2 K^{c\rho} \p^\mu F_{\mu\rho}
\label{zero1}
\ee
 It is straightforward to generalise this for
the covariantised angular momentum and one finds,
\be
\hat {\cal L}_{ab} \hat F^{abc} = 2s l^2\Big(1+s\frac{x^2}{4l^2}\Big)^2 K^{c\rho} D^\mu F_{\mu\rho}
\label{zc4}
\ee

Using the map (\ref{new9}) for the currents, the equation of motion on the A(dS) space finally gets
projected on the flat space as,
\be
\Big(1+s\frac{x^2}{4l^2}\Big)^2 K^{c\rho}\Big( D^\mu F_{\mu\rho} +\hat j_\rho\Big) = 0
\label{zc5}
\ee
This equation is now multiplied by the Killing vector $K_c^\lambda$. Using the identity among the Killing
vectors yields,
\be
\Big(1+s\frac{x^2}{4l^2}\Big)^4 \Big( D^\mu F_{\mu\lambda} +\hat j_\lambda\Big) = 0
\label{zc6}
\ee
 
The zero curvature limit $(l\rightarrow \infty)$ is now taken. The prefactor simplifies to unity and the standard flat
space Yang Mills equation
with sources is reproduced.

\section{Conclusions}

\bigskip

We have provided a manifestly  covariant formulation of non-abelian interacting gauge theories defined on the A(dS) hyperboloid. The various expressions, at each step of the analysis, preserved this covariance under the appropriate kinematic groups associated with the A(dS) space. A distinctive feature was to bypass the general formulation of field theories defined on a curved space \cite{wald} in favour of exploiting the symmetry properties peculiar to the A(dS) hyperboloid. This enabled us to set up a formulation that was general enough to include both de Sitter as well as anti de Sitter space times, arbitrary dimensions, non-abelian gauge groups and higher rank tensor fields. Also, a complete one to one mapping with the corresponding results on a flat Minkowski space-time was established. Using this correspondence it was possible to show that in the zero cuvature limit, the A(dS) field equation passed on to the flat space field equation. This was reassuring since it is known that the groups of the A(dS) space are deformations of the Poincare group which is the kinematical group of flat Minkowiki space.

     Our method consists in embedding the d-dimensional A(dS) space in a flat (d+1) dimensional space, sometimes called the ambient space. While this is a time honoured approach \cite{gutz,deserf}, we have deviated on two important issues. First, instead of working with arbitrary coordinate transformations that provide the map between the A(dS) space and the flat space, a particular type - the stereographic projection - has been used. An advantage of this is that this projection being conformal, the coordinate transformations were expressed in terms of the conformal Killing vectors. These vectors were explicitly computed by solving the Cartan-Killing equation. All expressions were thereby written in terms of the Killing vectors. Various properties of these vectors derived here were used to obtain the results compactly and transparently. The second distinctive feature was to avoid group theoretical techniques based on Casimir operators to construct the lagrangian or the action. We gave maps, involving the Killing vectors, for projecting the gauge fields on the flat space to the A(dS) space. Using these maps the action on the A(dS) space was constructed from a knowledge of the flat space action. The action so obtained was manifestly covariant under the kinematical symmetry group of the A(dS) hyperboloid. All derivatives appeared only through the angular momentum operator $L_{ab}$. In other approaches, apart from $L_{ab}$, the usual derivative $\partial_a$ also appears. This has to be removed by choosing subsidiary conditions which are obviously not required here. Our results are in general valid for arbitrary dimensions. In particular, the expression for the axial anomaly in A(dS) space was given for any $D=2n$ dimensions.

       We analysed the Yang-Mills theory and the two form gauge theory in details. Extension to higher rank tensor fields is straightforward. Also, a discussion of the matter (fermionic) sector was provided. The Lorentz gauge fixing condition was analysed. It appeared in two equivalent versions, one of which did not have any free index (reminiscent of the usual Lorentz gauge on a flat space) while the other had a single free index. The utility of both forms was   revealed. Specifically, the equivalence of our abelian gauge field equation with that obtained in other (say ambient) formalisms (after imposing appropriate subsidiary conditions) was established in the Lorentz gauge fixed sector, where  both versions of the gauge fixing had to be employed.

       We have computed the singlet (axial) anomaly as well as the non-abelian covariant and consistent chiral anomalies. This was done by projecting the relevant expressions from the flat to the A(dS) space. For the singlet case, we also computed the redefined expression for the axial current such that it was anomaly free. However, the current was no longer gauge invariant. This revealed the interplay between the anomaly and gauge invariance, exactly as happens for the flat example. In the non-abelian context the counterterm connecting the covariant and consistent anomalies has been calculated.

      The dual field tensor was introduced from which a form of the Bianchi identity was given. Electric-magnetic duality rotations swapping this identity with the equations of motion were found.

      We feel our approach gives an intuitive understanding of the closeness of formulating gauge field theories on flat and A(dS) space-time. Apart from the issues dealt here, a further application would be to develop the complete BRST formulation. This was earlier done by one of us \cite{banerjee1} in a collaborative work for the case of the hypersphere (an n-dimansional sphere embedded in (n+1)-dimensional flat space). Also, a possible connection between massless and massive higher rank tensor theories with superstring theory could be envisaged.

\section{Appendix: Variation Principle and Boundary Condition}

\bigskip

   The equation of motion of any system can be derived by using the variation principle according to which the action of the system is extremised. Here we will explicitly show how the equation of motion (\ref{field10}) comes by extremising the corresponding action (\ref{v21}) with suitable boundary conditions. The compatibility of these conditions on the A(dS) space and the flat space is also shown.

     Varying the action (\ref{v21}) and using the definition for field tensor (\ref{v12}) we obtain,
\ber
\delta S&=&-\frac{s}{6l^2}\int d\Omega Tr[\delta \hat{F}_{abc}\hat{F}^{abc}]
\nonumber
\\
&=&-\frac{s}{2l^2}\int d\Omega Tr[\delta(L_{ab}\hat{A}_c-ir_a[\hat{A}_b,\hat{A}_c])\hat{F}^{abc}]
\nonumber
\\
&=&-\frac{s}{2l^2}\int d\Omega Tr[L_{ab}\delta\hat{A}_c.\hat{F}^{abc}-2ir_a[\hat\delta{A}_b,\hat{A}_c]\hat{F}^{abc}]
\label{app1}
\eer
where in the last line we have used the antisymmetric property of $\hat{F}_{abc}$.
Now,
\ber
\int d\Omega (L_{ab}\delta\hat{A}_c).\hat{F}^{abc}
&=&\int d\Omega [(r_a\partial_b-r_b\partial_a)\delta\hat{A}_c].\hat{F}^{abc}
\nonumber
\\
&=&2\int d\Omega (\partial_b\delta\hat{A}_c)r_a\hat{F}^{abc}
\label{app2}
\eer
Using the expression for invariant measure (\ref{invmea}) we obtain,
\ber
&&\int d\Omega (L_{ab}\delta\hat{A}_c).\hat{F}^{abc}
=2l\int \frac{d^4r}{r_4}[(\partial_b\delta\hat{A}_c)r_a\hat{F}^{abc}]
\nonumber
\\
&=&2l\int \frac{d^4r}{r_4}(\eta{_b}{^\mu}\partial_\mu\delta\hat{A}_c)r_a\hat{F}^{abc}
\nonumber
\\
&=&2l\int d^4r(\partial_\mu\delta\hat{A}_c)\frac{1}{r_4}r_a\hat{F}^{a\mu c}
\nonumber
\\
&=&2l\int d^4r[\partial_\mu(\delta\hat{A}_c\frac{1}{r_4}r_a\hat{F}^{a\mu c})]-2l\int d^4r[\delta\hat{A}_c\partial_\mu(\frac{1}{r_4}r_a\hat{F}^{a\mu c})]
\label{app3}
\eer
Here $\partial_\mu = \frac{\partial}{\partial{r^\mu}}$. By Gauss' divergence theorem the first term of the above gives the surface term which vanishes at the boundary if we consider $\delta\hat{A}_c=0$ at the boundary. Using this boundary condition the above expression simplifies to,
\be
\int d\Omega (L_{ab}\delta\hat{A}_c).\hat{F}^{abc}=-2l\int d^4r \delta\hat{A}_c\partial_\mu[\frac{1}{r_4}r_a\hat{F}^{a\mu c}]
\label{app4}
\ee
Now,
\ber
&&\partial_\mu[\frac{1}{r_4}r_a\hat{F}^{a\mu c}]
=-\frac{1}{(r_4)^2}\frac{\partial r_4}{\partial r^\mu}r_a\hat{F}^{a\mu c}+\frac{1}{r_4}\partial_\mu[r_\nu\hat{F}^{\nu\mu c}+r_4\hat{F}^{4\mu c}]
\nonumber
\\
&=&-\frac{1}{(r_4)^2}\frac{\partial r_4}{\partial r^\mu}r_a\hat{F}^{a\mu c}+\frac{1}{r_4}[\hat{F}^{\mu\mu c}+r_\nu\partial_\mu\hat{F}^{\nu\mu c}+\partial_\mu r_4.\hat{F}^{4\mu c}+r_4\partial_\mu\hat{F}^{4\mu c}]
\label{app5}
\eer
Since $\hat{F}^{\mu\mu c}=0$ and $\frac{\partial r_4}{\partial r^\mu}=-\frac{sr_\mu}{r_4}$ the above expression reduces to,
\ber
&&\partial_\mu[\frac{1}{r_4}r_a\hat{F}^{a\mu c}]
\nonumber
\\
&=&\frac{s}{(r_4)^3}r_\mu(r_\nu\hat{F}^{\nu\mu c}+r_4\hat{F}^{4\mu c})+\frac{1}{r_4}(r_a\partial_\mu\hat{F}^{a\mu c}-\frac{sr_\mu}{r_4}\hat{F}^{4\mu c})
\nonumber
\\
&=&\frac{s}{(r_4)^3}r_\mu r_\nu\hat{F}^{\nu\mu c}+\frac{s}{(r_4)^2}r_\mu\hat{F}^{4\mu c}+\frac{1}{r_4}r_a\partial_\mu\hat{F}^{a\mu c}-\frac{s}{(r_4)^2}r_\mu\hat{F}^{4\mu c}
\label{app6}
\eer
The first term vanishes as $\hat{F}^{\nu\mu c}$ is antisymmetric in $\mu$ and $\nu$ and second and last terms cancel each other. Hence we get,
\ber
\partial_\mu[\frac{1}{r_4}r_a\hat{F}^{a\mu c}]&=&\frac{1}{r_4}r_a\partial_\mu\hat{F}^{a\mu c}
\nonumber
\\
&=&\frac{1}{r_4}r_a\partial_b\hat{F}^{abc}
\nonumber
\\
&=&\frac{1}{2r_4}L_{ab}\hat{F}^{abc}
\label{appr}
\eer
Substituting this in (\ref{app4}) we obtain,
\be
\int d\Omega(L_{ab}\delta\hat{A}_c)\hat{F}^{abc}=-\int d\Omega \delta\hat{A}_cL_{ab}\hat{F}^{abc}
\label{app8}
\ee
Likewise the second term of (\ref{app1}) can be simplified as,
\ber
\int d\Omega Tr[-2r_a[\delta\hat{A}_b,\hat{A}_c]\hat{F}^{abc}]
&=&\int d\Omega Tr[-2r_a(\delta\hat{A}_b\hat{A}_c-\hat{A}_c\delta\hat{A}_b)\hat{F}^{abc}]
\nonumber
\\
&=&\int d\Omega Tr (-2r_a\delta\hat{A}_b[\hat{A}_c,\hat{F}^{abc}])
\nonumber
\\
&=&\int d\Omega Tr(\delta\hat{A}_c.2r_a[\hat{A}_b,\hat{F}^{abc}])
\nonumber
\\
&=&\int d\Omega Tr\{\delta\hat{A}_c(r_a[\hat{A}_b,\hat{F}^{abc}]-r_b[\hat{A}_a,\hat{F}^{abc}])\}
\label{app9}
\eer
Substituting (\ref{app8}) and (\ref{app9}) in (\ref{app1}) we obtain,
\ber
\delta S=\frac{s}{2l^2}\int d\Omega Tr[\delta\hat{A}_c\{L_{ab}\hat{F}^{abc}-ir_a[\hat{A}_b,\hat{F}^{abc}]+ir_b[\hat{A}_a,\hat{F}^{abc}]\}]
\label{app10}
\eer
Hence $\delta S=0$ (subjected to the boundary condition $\delta\hat{A}_a=0$) yields, 
\be
\{L_{ab}-i[r_a\hat{A}_b-r_b\hat{A}_a,]\}\hat{F}^{abc}=0
\ee
i.e
\be
{\hat{\cal{L}}}_{ab}\hat{F}^{abc}=0
\ee
which is the equation of motion (\ref{field10}).

    The equivalence of this equation of motion with the stereographically projected YM equation of motion on flat space was shown in section-8.  Now the equation of motion on flat space is derived by the variation principle where we use the boundary condition, $\delta A_\mu=0$. Again, the vector field on the A(dS) space is the projection of the flat space field through the Killing vector which we have shown in (\ref{k}). Taking the variation of (\ref{k}) we find, $\delta\hat{A}_a=K{_a}{^\mu}\delta A_\mu=0$ on the boundary which is precisely the boundary condition used in the obtention of the equation of motion on A(dS) space. Therefore  we can conclude that the boundary condition on the flat space is really compatible with the boundary condition on the A(dS) space as the fields and their variations are connected by a Killing vector.

\end{document}